\documentclass{aa}  

\usepackage{graphicx}
\usepackage{txfonts}
\usepackage{lipsum}
\usepackage{hyperref}
\usepackage{subcaption}    
\usepackage{cuted}
\usepackage{lscape}             % to rotate a single page table, example in appendix.
\usepackage{placeins}           % useful with \FloatBarrier, to keep 
\usepackage[usenames]{color}
\usepackage{xcolor, colortbl}

\begin{document}

%%%%%%%%%%%%%%%%%%%%%%%%%%%%%%%%%%%%%%%%
% if you use custom commands in your title,
% ensure to check your title when submitting!
%%%%%%%%%%%%%%%%%%%%%%%%%%%%%%%%%%%%%%%%
   \title{Tracing Galaxy Bias Through the Cosmic Web: The Role of Filaments}

   %\subtitle{Subtitle}

%%%%%%%%%%%%%%%%%%%%%%%%%%%%%%%%%%%%%%%%
% Please separate each author with the \and command
%
% Please do not include ORCIDs next to author names.
% Only ORCIDs authenticated by individual authors in EDPS
% editorial system will be taken into account.
% ORCIDs included here will be removed.
%%%%%%%%%%%%%%%%%%%%%%%%%%%%%%%%%%%%%%%%

   \author{Constanza A. Soto-Suárez\inst{1}\fnmsep\inst{2}\fnmsep\thanks{Corresponding author: constanza.sotos@usm.cl}
        \and Antonio D. Montero-Dorta\inst{3}
        \and Daniela Galárraga-Espinosa\inst{4}
        \and Ignacio G. Alfaro\inst{3} \and Andrés Balaguera-Antolínez\inst{5} \and Ravi K. Sheth\inst{6,7} \and Pablo López\inst{8,9}
        }

   \authorrunning{C. A. Soto-Suárez et al.}
   
   \institute{Departamento de Física, Universidad Técnica Federico Santa María, Casilla 110-V, Avda. España 1680, Valparaíso, Chile 
   \and Instituto de Física, Pontificia Universidad Católica de Valparaíso, Casilla 4950, Valparaíso, Chile
   \and Departamento de Física, Universidad Técnica Federico Santa María, Avenida Vicuña Mackenna 3939, San Joaquín, Santiago, Chile
   \and Kavli IPMU (WPI), UTIAS, The University of Tokyo, Kashiwa, Chiba 277-8583, Japan
   \and Independent Researcher
   \and Center for Particle Cosmology, University of Pennsylvania, Philadelphia, PA 19104, USA 
   \and The Abdus Salam International Center for Theoretical Physics, Strada Costiera 11, Trieste 34151, Italy
   \and Observatorio Astronómico de Córdoba, Universidad Nacional de Córdoba (UNC), Francisco N. Laprida 854, Córdoba, Argentina
   \and
   Instituto de Astronomía Teórica y Experimental, CONICET-UNC, Laprida 922, Córdoba, Argentina\\
   }

   \date{Received --}

% \abstract{}{}{}{}{}
% 5 {} token are mandatory
 
  \abstract
   {The large-scale clustering of galaxies depends not only on their internal properties but also on their location within the cosmic web, which defines the anisotropic environments where galaxies form and evolve. Filaments serve as bridges through which galaxies and dark matter flow from low-density regions toward the highest-density nodes. A key aspect to investigate in this context is the galaxy bias within the cosmic web, with emphasis on cosmic filaments, which quantifies how these tracers follow the underlying dark matter density field in different environments.}
   {We aim to map galaxy bias through the cosmic web, with a special emphasis on filaments. We look for dependencies on filament properties, such as length and density, and characterize the spatial variations of the large-scale bias along the filamentary spine.} 
   {We applied the DisPerSE algorithm to identify the cosmic web in the TNG300 volume of the IllustrisTNG simulation. To measure large-scale galaxy bias, we utilized an object-by-object estimator, which provides advantages over standard estimators.}
   {We find that galaxies in node outskirts exhibit the highest large-scale bias values, reaching up to $\sim4$ times the values expected from theoretical models based on halo mass alone. Low-mass red galaxies in filaments and filament outskirts also display enhanced bias, which is strongly reduced after excluding galaxies close to nodes. Together, these results suggest that proximity to massive nodes plays a central role in shaping environmental secondary bias. Furthermore, galaxy bias decreases with filament length, from mean values of $1.4$ for short filaments to $-0.5$ for longer structures. This relation persists at fixed halo mass and galaxy color, and is not primarily driven by the average local galaxy density of the filament. Finally, short filaments exhibit an approximately uniform longitudinal bias profile, whereas long filaments show an increase in normalized bias from $\sim0.85$ near the saddle point to $\sim1.06$ close to the node.}
   {}

   \keywords{large-scale structure of the Universe --
                Galaxies: formation, statistics --
                Methods: numerical, statistics
               }

   \maketitle
   \nolinenumbers

\section{Introduction}
\label{sec:intro}

To first approximation, the Universe is homogeneous and isotropic on sufficiently large scales. However, small matter inhomogeneities, seeded by quantum fluctuations in the early Universe, grow under gravity into the large-scale structure (LSS) we observe today \citep[e.g.,][]{Zeldovich1970, Peebles1980}.  One of the main goals of LSS studies is to understand how the matter distribution develops into structures such as nodes, filaments, walls, and voids (\citealt{Shen2006, Hahn2007, van_de_Weygaert2011, cautun2014, Kitaura2021, Peebles1980, Aycoberry2024}), ultimately giving rise to the cosmic web (\citealt{Bond1996}).

Within this framework, filaments act as the structural "highways" of the Universe, serving as bridges of matter through which galaxies and dark matter flow from low-density regions toward the highest-density nodes \citep{Haarlem1993}. For this reason, the study of cosmic filaments has become a fundamental topic in modern astronomy and cosmology. Recent numerical simulations suggest that filaments contain approximately 50\% of the total mass of the Universe \citep{cautun2014, Ganeshaiah2019}. Consequently, they play a key role in models of galaxy formation and evolution, providing the environments where galaxies undergo "pre-processing", a series of physical transformations that occur before they enter the dense environments of galaxy clusters \citep[e.g.,][]{Martinez2016, Sarron2019, GalarragaEspinosa2023, OKane2024, Aguerri2026}. Filaments are thought to influence a wide range of galaxy properties, including the alignment of their angular momentum (spin) with the surrounding LSS \citep[e.g.,][]{Pereyra2020, Barsanti2022, tobar2026}, the suppression of star formation activity, and the enhancement of stellar mass growth \citep[e.g.,][]{Malavasi2017, Kraljic2018, Rost2020, OKane2024}; see also \citet{TojeiroKraljic2025} for a recent review. Furthermore, because their evolution is sensitive to the expansion history of the Universe, cosmic filaments are emerging as promising complementary probes of dark energy and alternative theories of gravity \citep[e.g.,][]{Llinares2014, Leonard2015, Cadiou2020, Lee2020}.

Galaxies act as observable tracers of the underlying matter distribution in the Universe. However, their spatial distribution does not perfectly follow that of the total matter field, since their clustering is shaped by both the properties of their host dark matter halos and the galaxies themselves \citep{Desjacques2018}. This connection between the galaxy and matter distributions is commonly described through the concept of galaxy bias, which quantifies the statistical relation between luminous tracers and the underlying dark matter density field \citep{Kaiser1984, Tegmark1998}. Characterizing galaxy bias is essential for accurately extracting cosmological information from LSS observations and for understanding the physical processes driving structure formation \citep{Mo1996, Hamaus2016, Singh2020}. On sufficiently large scales, the linear bias is primarily determined by the peak height of primordial density fluctuations and is therefore closely related to the mass of the host dark matter halo \citep{Mo1996, ShethTormen1999, Sheth2001}. However, at fixed halo mass, the clustering amplitude exhibits additional dependencies on halo properties such as concentration, spin, formation time, or the environment. This phenomenon is commonly referred to as secondary halo bias \citep{ShethTormen2004, Wechsler2006, Gao2007, 2018Salcedo, SatoPolito2019, Montero-Dorta2021, Contreras2021, tucci2021, Balaguera2024, monterodorta-rodriguez2024, MonteroDorta_Rodriguez2026}.

The importance of studying large-scale galaxy bias lies in its close connection to the processes that drive structure formation, particularly the dependence of halo assembly on the surrounding large-scale environment. Since cosmic filaments constitute the dominant channels through which matter is accreted onto halos and galaxies, they are expected to play a fundamental role in shaping these environmental dependencies. Crucially, the anisotropy of the cosmic web is a key driver of secondary (or assembly) bias \citep{Ramakrishnan2019}, as the large-scale tidal field can inhibit or facilitate halo growth depending on the geometry of the environment. Theoretical models have shown that mass accretion rates and formation times are sensitive to the orientation with respect to the distance from the surrounding cosmic web structures, suggesting that bias is not merely a function of local density but is deeply tied to the anisotropic tidal forces \citep{Sheth2013, Borzyszkowski2017, Musso2018}. Observational studies have also shown that large-scale galaxy bias depends on the anisotropy of the surrounding cosmic web environment \citep{ParanjapeSDSS2018, Alam2019}. More recently, galaxy bias has been shown to depend strongly on the distance to cosmic web structures, such as nodes, saddles, and filaments, revealing significant secondary bias signals that can exceed those linked to internal halo properties \citep{monterodorta-rodriguez2024}.

Characterizing how large-scale bias varies across the filamentary network can therefore provide valuable insight into the interplay between the growth of LSS and the formation and evolution of galaxies. In this work, we employ an object-by-object estimator of large-scale linear bias, taking advantage of several features that remain largely unexplored in traditional population-based approaches. By assigning a single bias value to each individual halo, bias itself becomes an intrinsic property that can be directly correlated with other halo characteristics and with the cosmic large-scale environment. This framework quantifies the contribution of each halo to the total linear bias of its population \citep{paranjape2018, Balaguera2024, montero2025b}. This enables a more detailed statistical characterization of how bias depends on both local and large-scale environments, providing a finer probe of the galaxy--halo connection and its relation to the cosmic web.

This work directly extends the recent analyses of \cite{montero2025b} and \cite{Alfaro2026}, who utilized this framework to characterize the environmental modulation of galaxy bias within cosmic voids. These previous studies established the existence of a robust `void bias profile', showing that individual bias correlates strongly with the distance to cosmic voids, efficiently capturing the underlying density variations in underdense regions. While those works focused on the most extreme low-density environments, the present study extends their analysis to high-density environments, focusing on the complex, anisotropic density field defined by the filamentary network.

Several methods have been developed to identify the cosmic web, including the Discrete Persistent Structure Extractor (\textsc{DisPerSE}; \citealt{Sousbie2011,Sousbieetal2011}), the \textsc{NEXUS/NEXUS+} algorithm \citep{Cautun2013}, the \textsc{Bisous} model \citep{Stoica2010, Tempel2016}, and the \textsc{T-Rex} method \citep{Bonnaire2020}. However, despite their widespread use, there is no unique or universally accepted criterion for defining the different components of the cosmic web. Different identification techniques, based on distinct assumptions regarding geometry, topology, velocity fields, or density thresholds, can produce substantially different catalogs even when applied to the same dataset \cite[see][for a discussion of filaments in particular]{Dhawalikar2024}. Despite these differences, comparison studies have shown that cosmic web finders recover broadly consistent statistical trends across the main environments, although the classification of individual objects can vary \citep[e.g.,][]{Libeskind2018}. Consequently, the inferred environmental dependence of galaxy properties may still depend on the adopted web definition \citep[e.g.,][]{Colberg2008, Alfaro2026, Rost2020}. In this work, we adopt the \textsc{DisPerSE} algorithm to identify the filamentary network in the hydrodynamical simulation IllustrisTNG\footnote{\url{http://www.tng-project.org}} \citep{Nelson2019} and investigate how individual galaxy bias depends on the properties and environments of cosmic filaments. It is also important to note that the filamentary network is intrinsically multiscale. Since our reconstruction is based on the galaxy distribution, our analysis is primarily sensitive to sufficiently prominent filaments that can be robustly traced by galaxies, while more tenuous structures may be underrepresented in our filament catalog \citep{Zakharova2023}.

\begin{figure}[h!]
    \centering
    \includegraphics[width=1\linewidth]{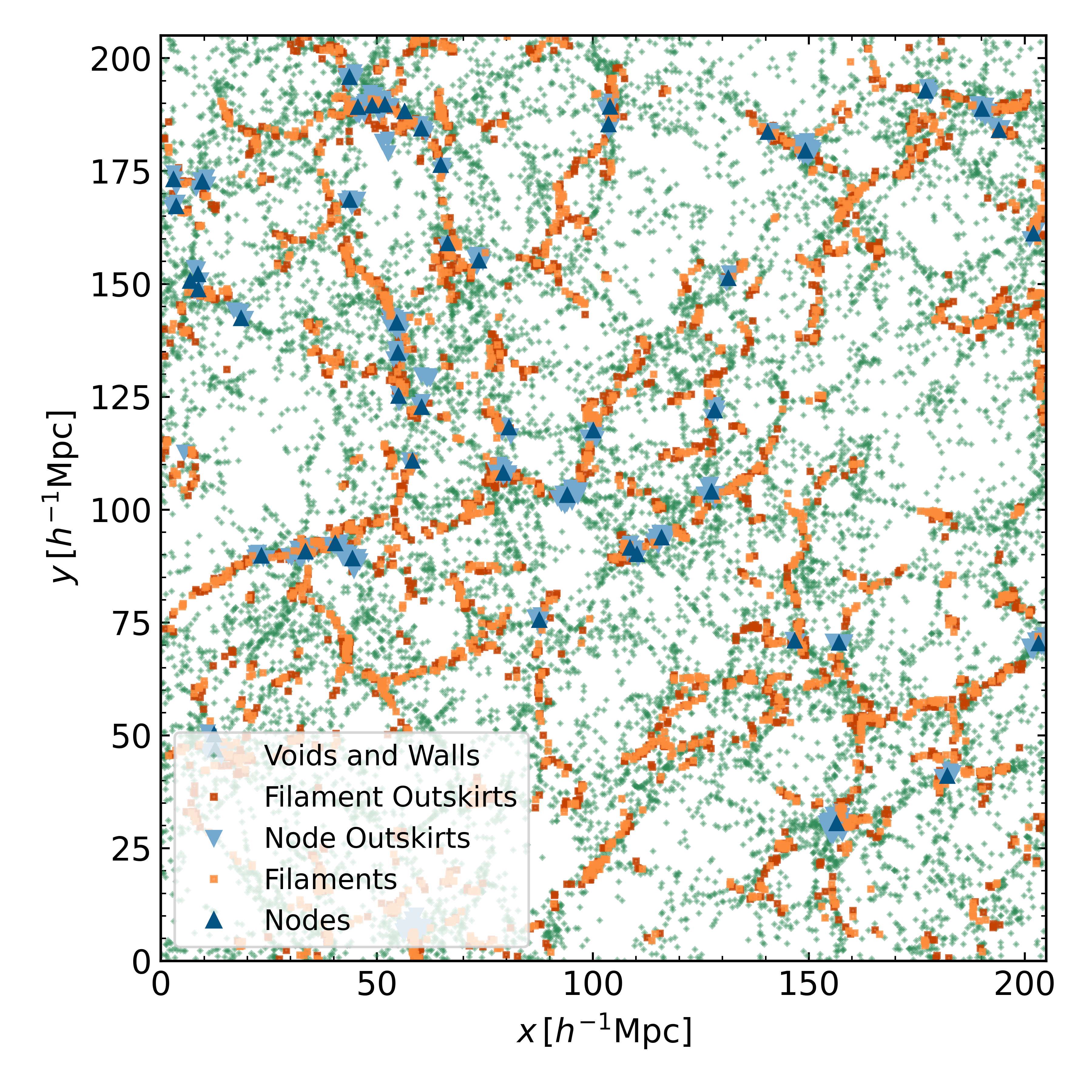}
    \caption{Visualization of the cosmic web classification in a $205\,h^{-1}{\rm Mpc} \times 205\,h^{-1}{\rm Mpc} \times 20\,h^{-1}{\rm Mpc}$ volume of TNG300 at $z=0$. Galaxies are colored according to their environment: nodes are shown as blue triangles, nodes outskirts as light-blue triangles, filaments as orange squares, filament outskirts as red squares, and voids/walls as green diamonds.}
    \label{fig:envs}
\end{figure}

The remainder of this work is organized as follows. In Section 2, we describe the galaxy samples and simulations used in this work, present the implementation of the cosmic web identification procedure, and introduce the methodology used to assign bias values to individual objects. Section 3 explores the overall distribution of these bias estimates, while Section 4 examines their dependence on filament properties and analyzes the longitudinal bias profiles along the filamentary skeleton. Section 5 provides an interpretation of our main results, which are summarized in Section 6.

%-----------------------------------------

\section{Data}
\label{sec:data}

For our analysis, we used data from the magneto-hydrodynamical cosmological simulation suite IllustrisTNG (hereafter TNG, for simplicity; \citealt{Pillepich2018b,Pillepich2018,Nelson2018_ColorBim,Nelson2019,Marinacci2018,Naiman2018,Springel2018}). The TNG simulations were performed using the moving-mesh code AREPO \citep{Springel2010}, which solves the equations of magnetohydrodynamics with a finite-volume scheme on a dynamic, unstructured Voronoi tessellation. They represent an improved version of the original Illustris simulations \citep{Vogelsberger2014a,Vogelsberger2014b,Genel2014}. The simulations incorporate sub-grid models for radiative metal cooling, star formation, chemical enrichment from Type II and Type Ia supernovae and AGB stars, and feedback from both stellar populations and supermassive black holes. These simulations were calibrated to reproduce several observational constraints, including the $z = 0$ stellar mass function, the cosmic star formation rate density, halo gas fractions, galaxy size distributions, and the black hole--galaxy mass scaling relation.

\begin{figure}[h!]
    \centering
    \includegraphics[width=1\linewidth]{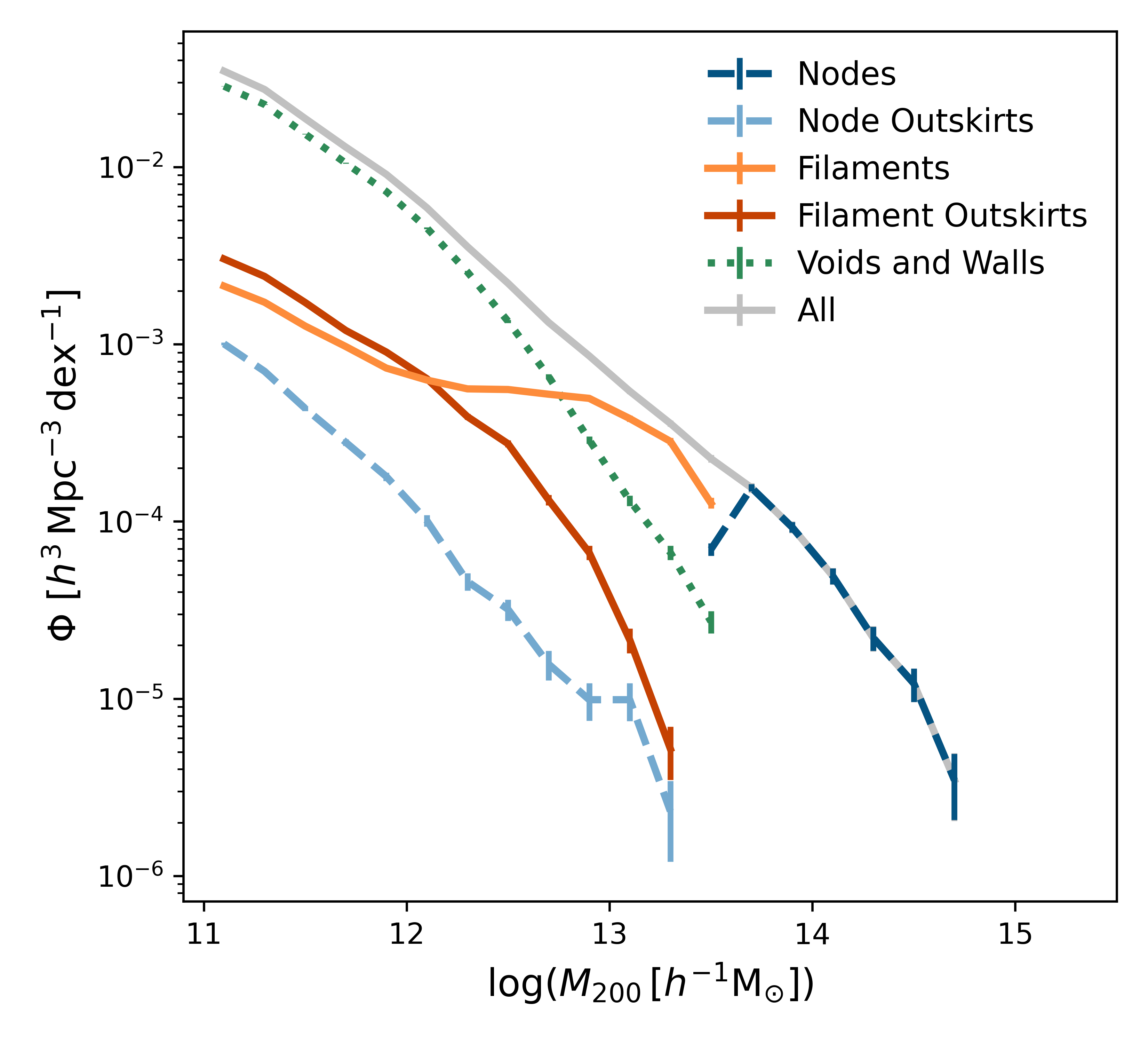}
    \caption{Halo mass function measured separately for each cosmic web environment at $z=0$. The gray solid line shows the halo mass function of the full galaxy sample, while the colored lines show the contribution from halos classified as nodes (blue), node outskirts (light-blue), filaments (orange), filament outskirts (red), and voids/walls (green). Error bars correspond to bootstrap standard errors estimated by resampling the halo catalog in each cosmic web environment.}
    \label{fig:hmf}
\end{figure}

Since this work focuses on the LSS of the Universe, we use the $z = 0$ snapshot of the TNG300-1 simulation (hereafter TNG300, for simplicity), which is the largest simulation box in the TNG suite. TNG300 adopts a periodic cubic box with a side length of 205 $h^{-1}$ Mpc. It follows the evolution of $2500^3$ dark matter particles, each with a mass of $4.0 \times 10^{7}$ $h^{-1}$ M$_{\odot}$, and $2500^3$ initial gas cells, each with a mass of $7.6 \times 10^{6}$ $h^{-1}$ M$_{\odot}$. The TNG300 simulation assumes a standard $\Lambda$ cold dark matter ($\Lambda$CDM) cosmology \citep{Planck2016}, with parameters $\Omega_{\rm m} = 0.3089$, $\Omega_{\rm b} = 0.0486$, $\Omega_\Lambda = 0.6911$, $H_0 = 100 \, h\, {\rm km \, s^{-1} Mpc^{-1}}$ with $h=0.6774$, $\sigma_8 = 0.8159$, and $n_s = 0.9667$.

We use the halo mass, $M_{\rm 200}$ [$h^{-1}{\rm M_{\odot}}$], defined as the total mass enclosed within a sphere of radius $R_{\rm 200}$, whose mean density is 200 times the critical density of the Universe. For galaxies, we compute the stellar mass, $M_*$ [$h^{-1}{\rm M_{\odot}}$], as the total mass of stellar particles associated with each subhalo. Central galaxies are identified as those subhalos whose index matches the \texttt{GroupFirstSub} entry of their corresponding FoF group in the TNG300 simulation. Our catalog includes all simulated central galaxies with $M_* > 10^{8.33}\,{h^{-1}\rm\, M_\odot}$ hosted by halos with $M_{200} \geq 10^{11}\,h^{-1}{\rm M_\odot}$, resulting in a total of $203\:844$. Galaxy colors are defined using the $g$ and $r$ photometric bands provided by the IllustrisTNG database \citep{Nelson2018_ColorBim}.

\subsection{Cosmic web classification}
\label{sec:cosmic_classification}

To identify the cosmic web components, we make use of the \textsc{DisPerSE} structure finder. \textsc{DisPerSE} is based on discrete Morse theory and assumes that the cosmic web can be adequately described within a mathematical framework known as the Morse complex, in which a set of manifolds can be associated with the different cosmic web structures. The method first computes the density field from a discrete tracer distribution, in this case galaxies, using the Delaunay Tessellation Field Estimator (DTFE; see \citealt{vandeWeygaert&Schaap2009}). It then identifies the critical points of the density field, namely maxima, minima, and saddle points, together with their associated manifolds, which define the different components of the cosmic web, corresponding to peaks, voids, walls, and
filamentary structures. Filaments are defined as sets of segments connecting maximum-density critical points, hereafter CPmax, to saddle points, following the ridges of the density field.

In this work, we follow a \textsc{DisPerSE} implementation similar to that of \citet{Galarraga2020} to identify filaments in TNG300. The robustness of this filamentary skeleton has been tested following the procedure introduced by \citet{Galarraga2024} and has been used in multiple studies of the cosmic web \citep[e.g.,][]{Rodriguez2025, Palma2026}. We use a $3\sigma$ persistence threshold on a smoothed DTFE density field computed from galaxies with $M_* \geq 10^{8.83}\,{h^{-1}\rm M_\odot}$. Following the approach introduced by \citet{GalarragaEspinosa2023}, we assign each galaxy to a single cosmic web environment among five categories: nodes, node outskirts, filaments, filament outskirts, and a combined voids-and-walls category. This classification is constructed hierarchically, ensuring that each galaxy is assigned to a single environment.

Galaxies in nodes are those residing within a sphere of radius $R_{200}$ centered on massive FoF halos with $M_{200} >   10^{13.53}\, h^{-1} \rm M_{\odot}$. Galaxies in node outskirts are those located within a spherical shell extending from $1\,R_{200}$ to $3\,R_{200}$ around these clusters. Galaxies in filaments are those lying within $0.68 \,h^{-1}\mathrm{Mpc}$ of a \textsc{DisPerSE} filament and not already classified as nodes or node outskirts, while galaxies in filament outskirts are those at distances between $0.68$ and $1.35\,h^{-1}\mathrm{Mpc}$ from a filament. Finally, galaxies not belonging to any of the previous categories are grouped into the voids and walls environment. To illustrate the resulting cosmic web classification, Figure \ref{fig:envs} shows a slice of thickness $20\;h^{-1}\mathrm{Mpc}$ containing the halos from the final galaxy sample described above. The galaxies are represented by symbols of different colors according to their environmental classification: green diamonds correspond to voids and walls, orange and red squares to filaments and filament outskirts, and blue and light-blue triangles to nodes and node outskirts. This figure provides a visual overview of the spatial distribution of the different environments identified by our classification scheme and highlights the connectivity of the filamentary network linking the densest structures. The number of galaxies and halos belonging to each environment is specified in Table \ref{tab:envs_catalogs} with their halo mass range.

\begin{table}
    \caption{Main characteristics of the final environmental catalog.}
    \label{tab:envs_catalogs}
    \centering
    {
    \scriptsize
    \begin{tabular}{c c c c c c}    
    \hline\hline 
    \textsc{DisPerSE} & $N_{\mathrm{gal}}$ & $M_{\mathrm{200c,\ min}}$ & $M_{\mathrm{200c,\ mean}}$ & $M_{\mathrm{200c,\ max}}$ \\
    \ classification & & [$h^{-1}M_{\odot}$] & [$h^{-1}M_{\odot}$] & [$h^{-1}M_{\odot}$] \\ 
    \hline\hline
    Voids and Walls & 161801 & 10$^{11.00}$ & 10$^{11.71}$ & 10$^{13.53}$ \\
    Nodes & 701 & 10$^{13.53}$ & 10$^{13.93}$ & 10$^{15.02}$ \\
    Node Outskirts & 4859 & 10$^{11.00}$ & 10$^{11.68}$ & 10$^{13.44}$ \\
    Filaments & 17883 & 10$^{11.00}$ & 10$^{12.41}$ & 10$^{13.53}$ \\
    Filament Outskirts & 18600 & 10$^{11.00}$ & 10$^{11.79}$ & 10$^{13.46}$ \\
    \hline
    \end{tabular}
    }
  \tablefoot{
    Column 1: Environmental classification; 
    Column 2: total number of galaxies;  
    Column 3: minimum value of halo masses; 
    Column 4: mean value of halo masses; 
    Column 5: maximum value of halo masses.    
  }   
\end{table}

To visualize how halo masses are distributed across the five cosmic web environments, Figure~\ref{fig:hmf} shows the halo mass function (HMF, in solid gray line for the full sample) in each category. As expected, massive halos are significantly less abundant than low mass halos across all environments. Nodes (blue dashed line) are dominated by the most massive halos, reflecting their role as the highest-density regions of the cosmic web. Their outskirts (light-blue dashed line), although still populated by relatively massive systems, show a noticeable decline at the high-mass end. In contrast, voids and walls (green dotted line) are primarily populated by low-mass halos and contain only a small fraction of the most massive objects. Filaments (orange solid line) occupy an intermediate regime, with a halo population that bridges the transition between dense nodes and underdense regions. A similar behavior is found for filament outskirts (red solid line), although with an even smaller contribution from the highest-mass halos.

\begin{figure}
    \centering
    \includegraphics[width=1.03\linewidth]{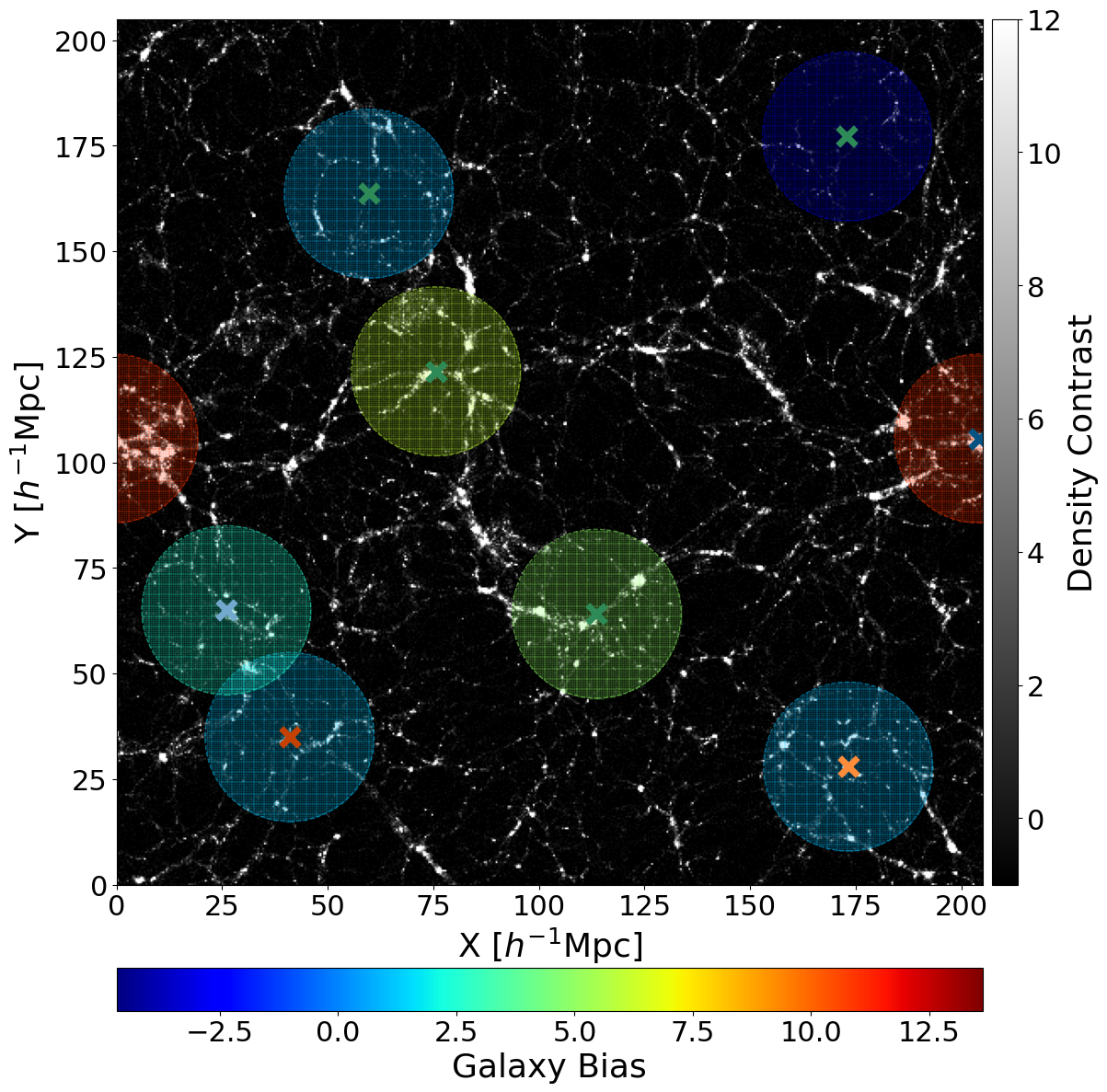}
    \caption{Illustration of the relation between individual galaxy bias and the surrounding large-scale environment. The background shows the projected density contrast field in a slice of the TNG300 simulation. Crosses mark selected galaxies, colored according to their cosmic web classification: nodes in blue, node outskirts in light-blue, filaments in orange, filament outskirts in red, and voids and walls in green. Around each selected galaxy, we show a circle of radius $20\,h^{-1}{\rm Mpc}$, colored according to the individual bias value of the corresponding galaxy.}
    \label{fig:bias_ilus}
\end{figure}

\subsection{Assigning individual galaxy bias}\label{sec:individual_bias}
\label{sec:ind_bias}

Regular estimators of halo bias and secondary halo bias are commonly based on ratios of (cross-)power spectra between halos and dark matter, or on two-point correlation functions, e.g., $b = \xi_{hm}/\xi_{mm}$, where $\xi_{mm}$ denotes the auto-correlation of the dark matter density field and $\xi_{hm}$ the cross-correlation between halos and the dark matter distribution itself. A limitation of these methods is that the bias is typically measured for subsets of halos, which can complicate analyses aimed at disentangling correlations among internal halo properties, environment, and bias. To overcome this limitation, we compute the large-scale linear bias using the object-by-object estimator introduced by \cite{paranjape2018}, also referred to as the \textit{individual bias}, $b_i$. Within this framework, the large-scale bias is interpreted as a sample mean in which each galaxy contributes through its own individual bias. 

This methodology is well suited to our analysis, since it facilitates the study of how galaxy bias behaves within anisotropic structures like cosmic filaments. By assigning an individual large-scale bias value to each galaxy, it allows us to characterize how galaxy bias varies with distance, to map longitudinal variations along the filamentary spine, and to explore dependencies on global filament properties such as length and density. Similar applications of the individual bias formalism can be found in \cite{Ramakrishnan2019, Contreras2021, Balaguera2024, balagueramontero2024, montero2025a, montero2025b, Alfaro2026}.

Following the prescription of \cite{Balaguera2024}, and employing basic properties of discrete Fourier transforms, the bias of a halo located at position $\mathbf{r}_i$ is given by

\begin{equation}
    b_i =
    \frac{\sum_{j,k_j<k_{\rm max}} N_k^j 
    \left\langle 
    e^{-i\mathbf{k}\cdot\mathbf{r}_i}
    \delta_{\mathrm{DM}}^{*}(\mathbf{k})
    \right\rangle_{k_j}}
    {\sum_{j,k_j<k_{\rm max}} N_k^j P_{\rm DM}(k_j)},
    \label{eq:bias}
\end{equation}
where $\delta_{\mathrm{DM}}(\mathbf{k})$ is the Fourier transform of the dark matter density field, $P_{\rm DM}(k_j)$ is the matter power spectrum, and $N_k^j$ is the number of Fourier modes in the $j$-th spherical shell. By construction, this estimator provides an isotropic measure of the large-scale bias, since the Fourier modes are averaged over spherical shells in $k$-space. The summation is performed over the range of wavenumbers for which the ratio between the halo and dark matter power spectra remains approximately constant, corresponding to the linear regime. In our analysis, we adopt $k_{\rm max}=0.2\,h\,\mathrm{Mpc}^{-1}$.  

Within this formalism, the effective large-scale bias of a population containing $N_G$ halos can be simply obtained as the mean of the individual biases,

\begin{equation}
    \langle b \rangle_G = \frac{1}{N_G}\sum_{i=1}^{N_G} b_i.
    \label{eq:bias_pop}
\end{equation}

It is worth noting that Eq.~(\ref{eq:bias}) should not be interpreted as describing a local connection between galaxies and dark matter. Instead, the individual bias is a proxy for the large-scale overdensity in the underlying matter field, as seen by each galaxy. This interpretation follows directly from the term $e^{-i\mathbf{k}\cdot\mathbf{r}i}\delta^*_{\rm DM}(\mathbf{k})$, where the phase factor $e^{-i\mathbf{k}\cdot\mathbf{r}_i}$ encodes the position of the galaxy within each Fourier mode of the dark matter density field. By summing over linear modes, the estimator therefore captures the large-scale matter environment associated with each tracer. In this sense, $b_i$ quantifies the large-scale environment surrounding each galaxy and corresponds to the contribution of that individual object to the large-scale bias of the population (Eq.~\ref{eq:bias_pop}).

\begin{figure}[h!]
    \centering
    \includegraphics[width=1\linewidth]{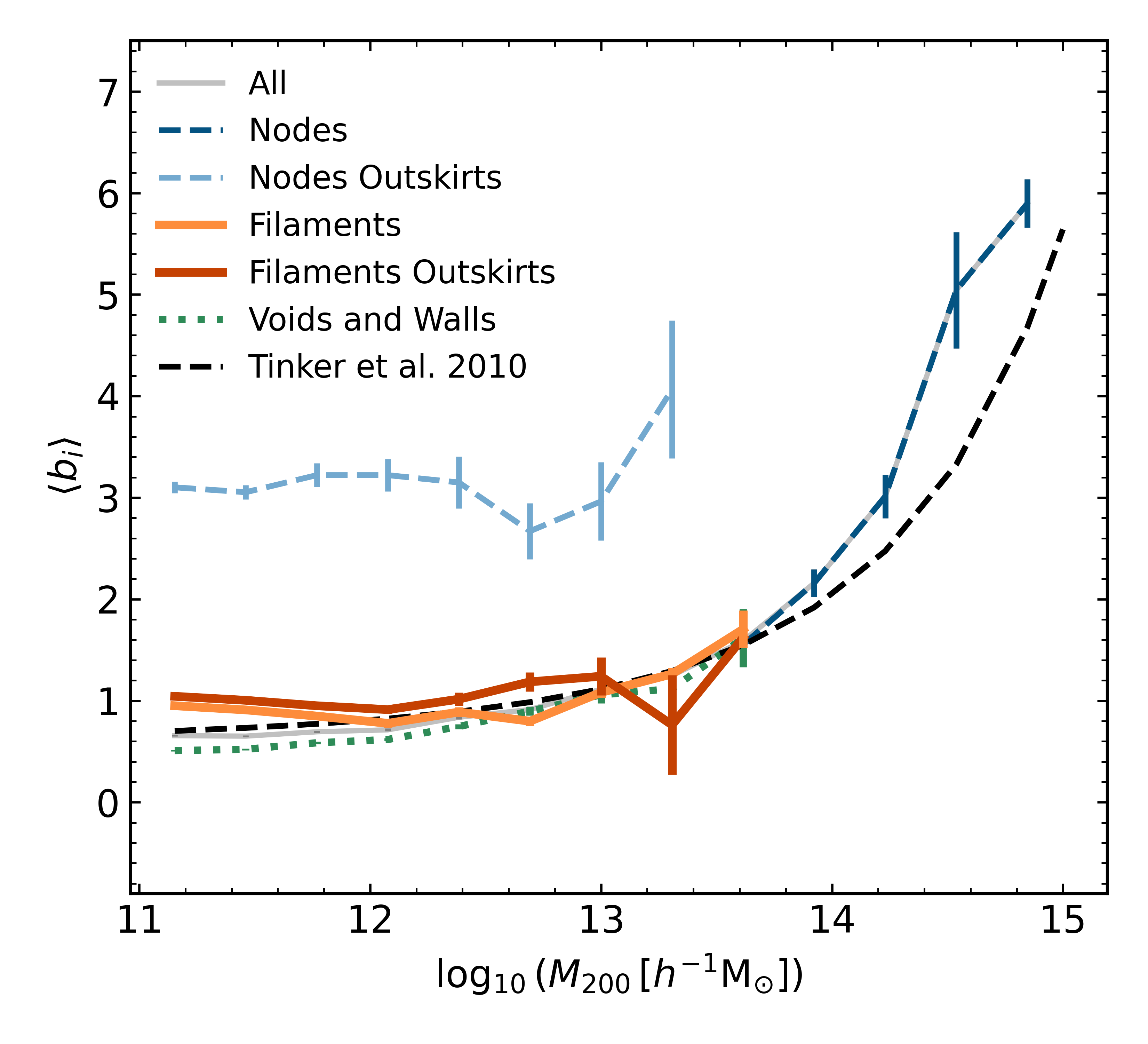}
    \caption{Mean individual galaxy bias, $\langle b_i\rangle$, as a function of halo mass for galaxies in the different cosmic web environments defined in Sect.~\ref{sec:cosmic_classification}. The gray solid line shows the full sample. The dark-blue and light-blue dashed lines correspond to nodes and node outskirts, respectively, while the orange and brown solid lines show filaments and filament outskirts. The green dotted line corresponds to galaxies in voids and walls. Error bars represent the standard error on the mean, estimated through bootstrap resampling. The black dashed line shows the fit of \citet{Tinker2010}.}
    \label{fig:biasfunc_all_cen_sat}
\end{figure}

This interpretation is illustrated in Figure~\ref{fig:bias_ilus}, where we show a set of randomly selected galaxies, colored according to the cosmic web classification defined in Sec.~\ref{sec:cosmic_classification}. Around each selected galaxy, we draw a circle of radius $20\,h^{-1}{\rm Mpc}$, colored according to its individual bias value. Galaxies located in denser environments, such as nodes and node outskirts, typically exhibit larger $b_i$ values, reflecting the enhanced large-scale matter overdensity around them. In contrast, galaxies associated with lower-density environments, such as voids and walls, generally display lower $b_i$ values and are embedded in large-scale regions with lower matter overdensities. The figure also shows that some galaxies assigned to lower-density cosmic web environments can still have high individual bias values when their surrounding large-scale matter distribution is influenced by nearby dense structures. This emphasizes that the individual bias is not determined solely by the local cosmic web classification, but by the underlying matter overdensity on the large scales probed by the estimator.

\section{The galaxy bias function in different cosmic environments}
\label{sec:bias_func}

In this section, we analyze the dependence of the individual contributions to the effective large-scale galaxy bias on halo virial mass, $M_{200}$, in different cosmic web environments, as defined in Sec.~\ref{sec:cosmic_classification}, and as a function of galaxy properties. We characterize this dependence through the bias function, defined here as the mean individual bias, $\langle b_i\rangle$, of galaxies within bins of virial halo mass, $M_{200}$. 

Figure~\ref{fig:biasfunc_all_cen_sat} shows the bias function for the full galaxy sample (gray solid line) and for the five cosmic web environments defined in Sec.~\ref{sec:cosmic_classification}. For comparison with standard bias estimates, which capture the expected increase of halo bias with halo mass, we also show the fitting formula of \citet{Tinker2010} as a dashed black line.

\begin{figure}[h!]
    \centering
    \includegraphics[width=1\linewidth]{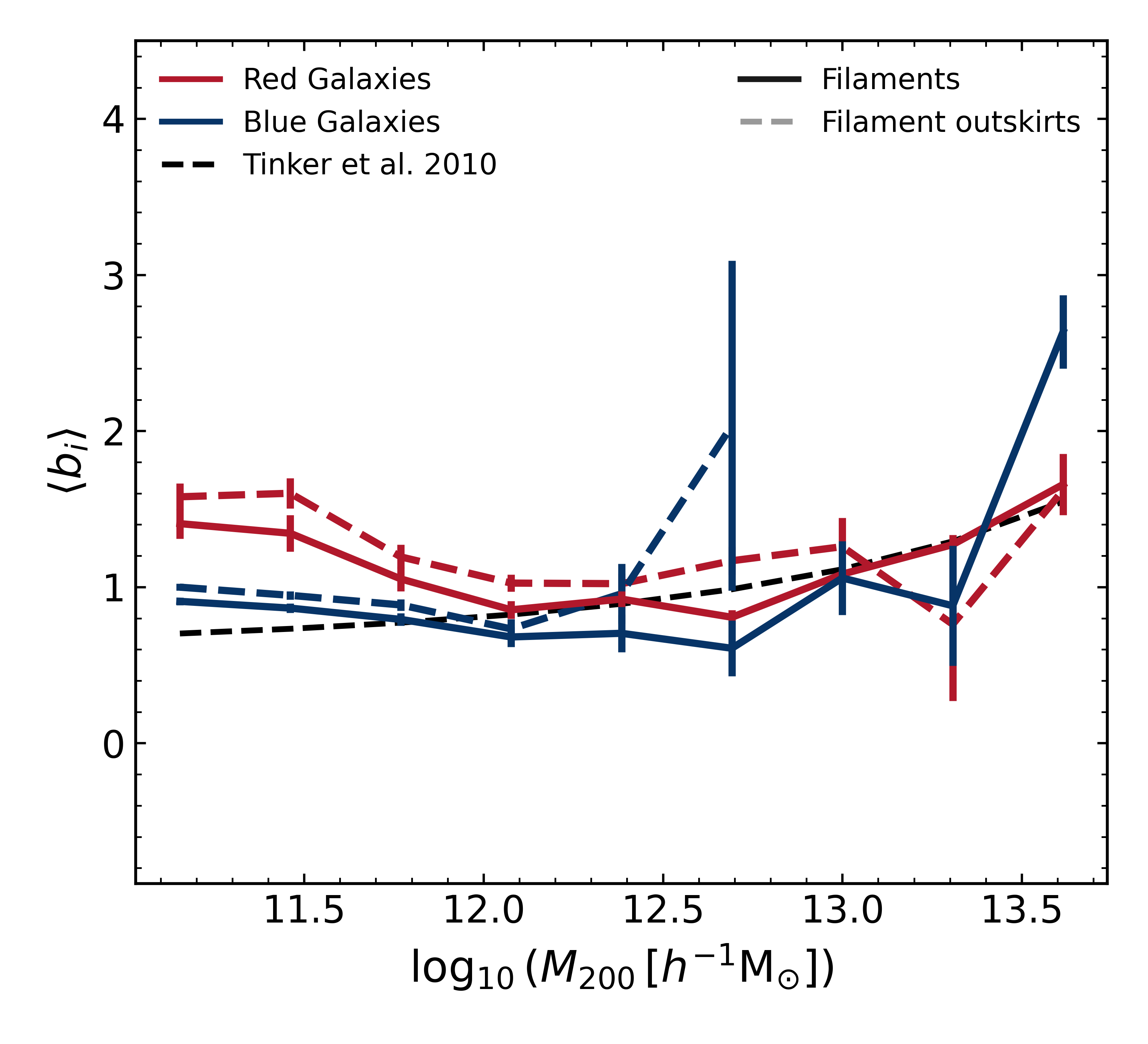}
    \caption{Bias function for red and blue galaxies in filaments and filament outskirts. Solid lines correspond to galaxies in filaments, while dashed lines correspond to galaxies in filament outskirts. Red and blue colors indicate red and blue galaxies, respectively. Error bars correspond to the standard error on the mean, derived from bootstrap resampling. The dashed black line shows the fit of \citet{Tinker2010}.}
    \label{fig:bf_colors}
\end{figure}

Overall, the bias function is positive in all cosmic web environments, and most of the measurements are in good agreement with the \citet{Tinker2010} fit. Nevertheless, clear environmental trends are observed in Fig.~\ref{fig:biasfunc_all_cen_sat}. At fixed halo mass, galaxies located in voids and walls (green dotted line) exhibit lower bias values than those in filaments (orange solid line) and filament outskirts (brown solid line), particularly for $M_{200} \leq 10^{12.5}\,h^{-1}{\rm M_\odot}$. In this mass range, the measured bias functions reach up to $\sim1.35$ and $\sim1.49$ times the \citet{Tinker2010} fit in filaments and filament outskirts, respectively, while galaxies in voids and walls exhibit bias values as low as $\sim0.71$ times the fit. This environmental dependence becomes even more pronounced in node outskirts (light-blue dashed line), where the bias function displays significantly higher values across the full mass range. In this environment, the mean bias reaches values of $4.1$, exceeding the value inferred from the \citet{Tinker2010} fitting formula by a factor of up to $\sim4$. This suggests that, at fixed halo mass, galaxies residing near the densest regions of the cosmic web exhibit higher large-scale bias than their counterparts in voids and walls, filaments, and filament outskirts. This enhancement near nodes highlights the important role of dense and massive environments in modulating the large-scale bias of galaxies and provides evidence for environmental secondary bias. Such an enhancement is consistent with recent findings in both simulations and observational data \citep[e.g.,][]{monterodorta-rodriguez2024, MonteroDorta_Rodriguez2026} and will be discussed in further detail in Section \ref{sec:disc}. At the high-mass end ($M_{200}\gtrsim10^{13.53}\,h^{-1}{\rm M_\odot}$), particularly galaxies in nodes, the departure from the \citet{Tinker2010} fit is likely affected by the limited statistics of massive halos in the TNG300 volume.

\begin{figure}[h!]
    \centering
    \includegraphics[width=1\linewidth]{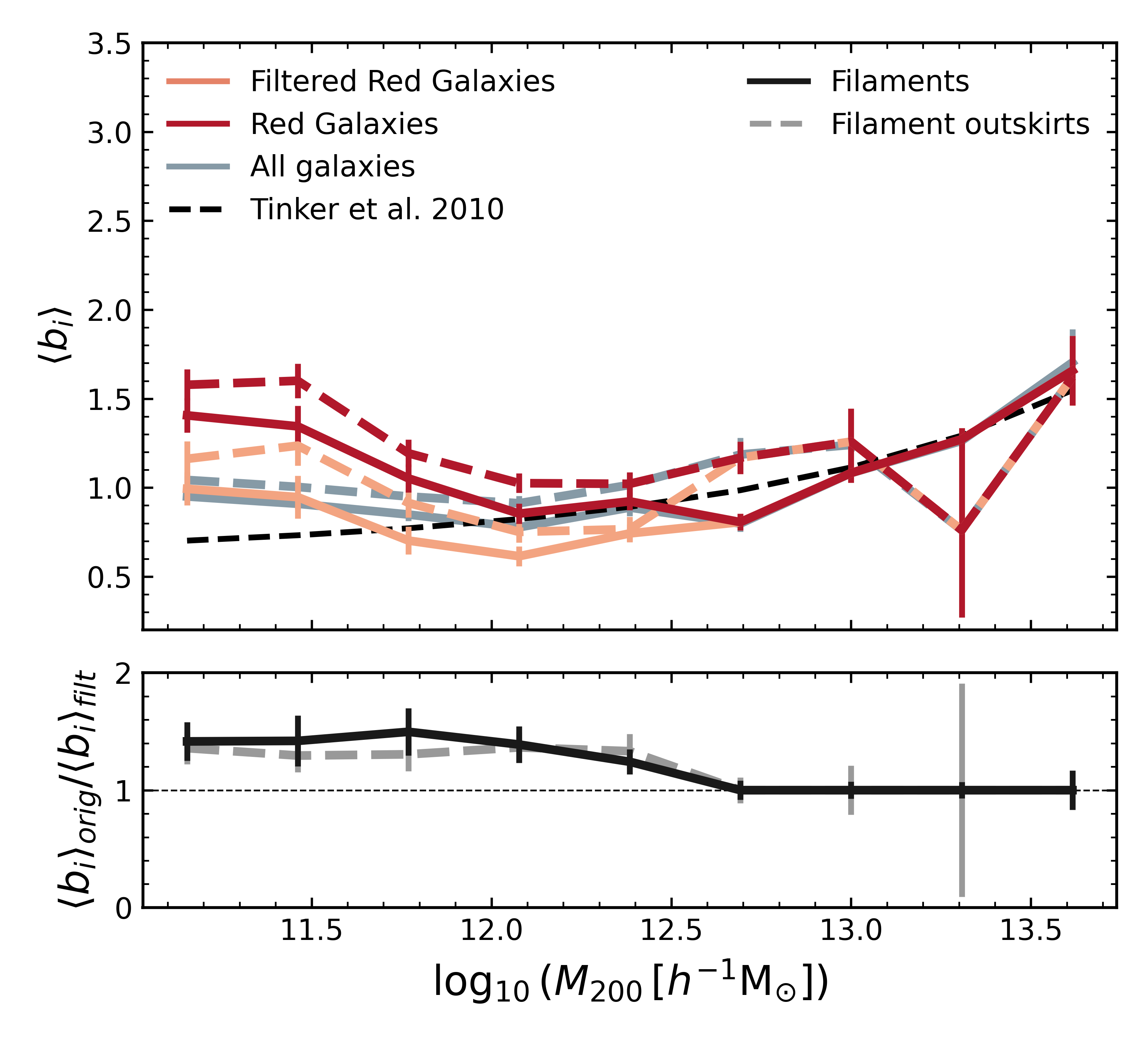}
    \caption{Top panel: Mean individual bias as a function of halo mass for galaxies in filaments (solid lines) and filament outskirts (dashed lines). Gray curves show the full galaxy sample, while red curves correspond to red galaxies. Pink curves display the same red-galaxy samples after excluding objects with $M_{200}<10^{12.5} \, h^{-1}{\rm M_\odot}$ located within $5\,h^{-1}{\rm Mpc}$ of a node. The black dashed line shows the linear halo bias fit of \cite{Tinker2010} model for reference. Bottom panel: Ratio between the original and filtered mean individual bias measurements for the red-galaxy samples $\langle b_i \rangle_{orig} / \langle b_i \rangle_{filt}$.}
    \label{fig:biasfilt_tng}
\end{figure}

We now focus on galaxies in filaments and filament outskirts. To explore how our bias estimation depends on galaxy properties, we divide this sample into red and blue populations. Following \citet{Lacerna2022}, we adopt a threshold of $g-r=0.6$. This threshold is also consistent with the color bimodality found in IllustrisTNG and has been used to separate red and blue galaxies in TNG300 \citep{Nelson2018_ColorBim}. Figure~\ref{fig:bf_colors} shows the resulting bias functions for the two populations. In both filaments (solid line) and filament outskirts (dashed line), red galaxies systematically exhibit higher mean bias values than blue galaxies at fixed halo mass over most of the halo mass range explored. A similar trend is found in the other cosmic-web environments, although here we focus on filaments and filament outskirts. This behavior is consistent with previous studies showing that, at fixed halo mass, systems hosting red galaxies are more strongly clustered than those hosting blue galaxies (e.g., \citealt{Wang_2008}), as well as with evidence for galactic conformity and environmental correlations extending beyond the virial radius (e.g., \citealt{Lacerna2022}). However, at the low-mass end, we identify a noticeable excess in the bias function of the red galaxy population. For $M_{200} \leq 10^{12.5}\;h^{-1}{\rm M_\odot}$, red galaxies exhibit systematically higher bias values than blue galaxies, by factors of $\sim 1.5$ and $\sim 1.6$ in filaments and filament outskirts, respectively. 

To investigate whether the low-mass upturn is driven by the proximity to dense environments, we recompute the bias--mass relation after removing galaxies with $M_{200}<10^{12.5}\,h^{-1}{\rm M_\odot}$ located within $5 \, h^{-1}{\rm Mpc}$ of a node. Figure~\ref{fig:biasfilt_tng} compares the original (red lines) and filtered samples (pink lines) for galaxies in filaments and filament outskirts. The effect of this selection is similar for galaxies inside filaments (solid lines) and galaxies in filament outskirts (dashed lines), indicating that their bias distribution is sensitive to the presence of nearby nodes. After the filtering procedure, the bias-mass relation in red galaxies from both environments becomes nearly indistinguishable from that measured in the complete galaxy samples (gray lines) of filaments and filament outskirts. In the low-mass range, the bottom panel shows that the original red-galaxy sample has higher bias values than the filtered sample by factors of up to $\sim1.5$ in filaments and $\sim1.4$ in filament outskirts. This result suggests that the excess in the bias signal of the low-mass red galaxy population is primarily driven by galaxies located close to massive nodes, rather than being an intrinsic property of the environments themselves.

\begin{figure}[h!]
    \centering
    \includegraphics[width=1\linewidth]{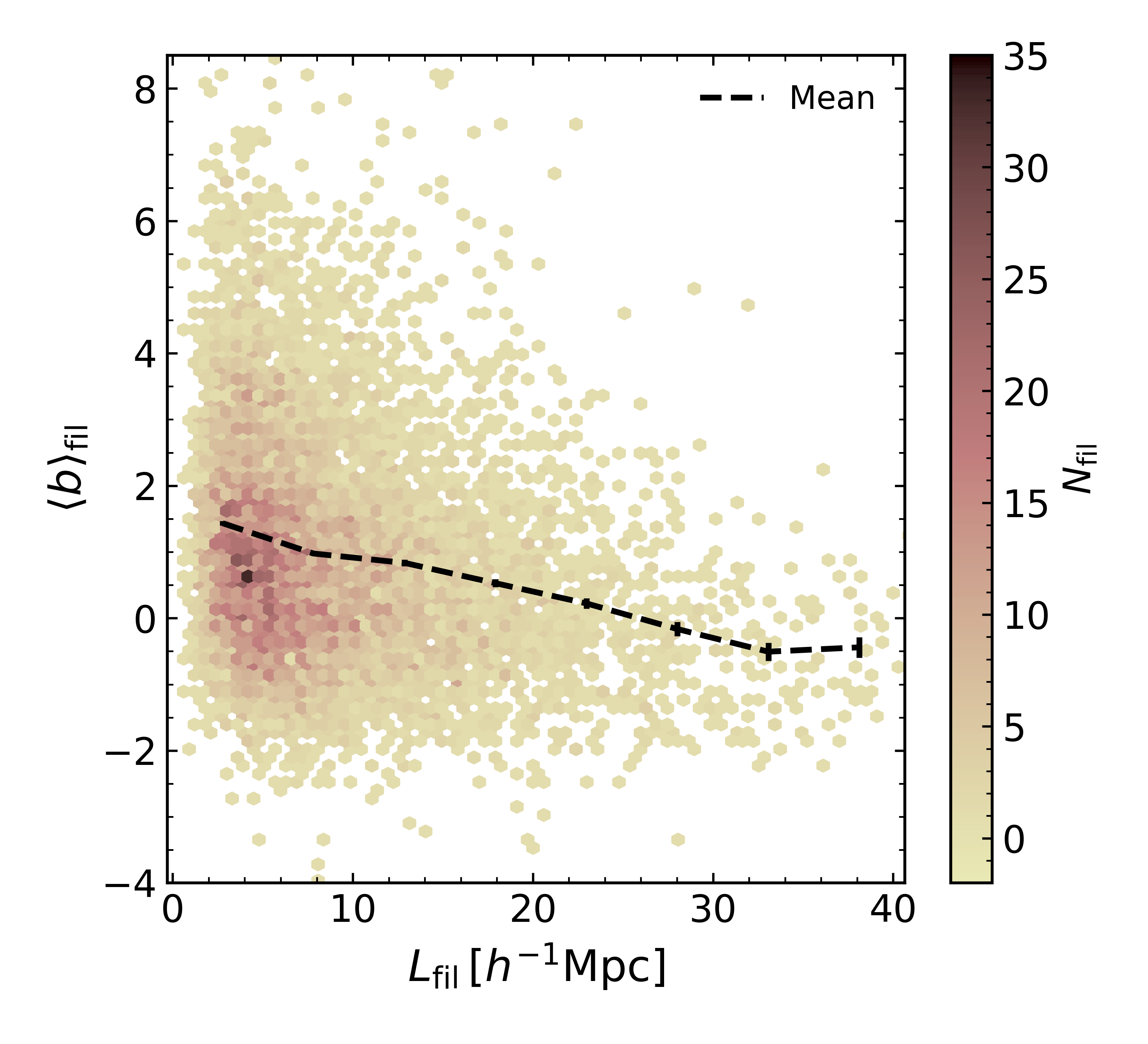}
    \caption{Mean galaxy bias of individual filaments as a function of filament length. Each point represents a single filament, colored by the number of galaxies associated with it. The black dashed curve shows the mean trend. Longer filaments exhibit systematically lower mean galaxy bias values.}
    \label{fig:mean_bias_fil}
\end{figure}

\section{Dissecting galaxy bias as a function of filament properties}
\label{sec:galaxies_in_filaments}

Motivated by the results of the previous section, the remainder of our analysis focuses primarily on galaxies residing in filaments and filament outskirts. Unlike nodes, which are compact structures characterized by uniformly high large-scale densities, filaments extend over tens of megaparsecs and span a much broader range of large-scale matter densities. Since the individual bias is sensitive to the matter distribution smoothed on large scales, only modest variations are expected within the relatively confined volume of nodes. In contrast, filaments provide a richer setting in which galaxies may reside in regions with significantly different large-scale matter densities while still belonging to the same cosmic web component (see Appendix \ref{app:bias_fil}). Furthermore, their intrinsically anisotropic nature, diverse morphologies, and role as the main channels connecting nodes make them particularly well suited for investigating how LSS influences the galaxy bias beyond the information encoded in halo mass alone.

\subsection{Galaxy bias as a function of filament length}

To perform the analysis focused on filamentary environments, we select galaxies classified as belonging to filaments or filament outskirts. For each of these galaxies, we assign a unique filament by selecting the closest \textsc{DisPerSE} filament segment. This procedure ensures that each galaxy is associated with a single filament, avoiding multiple assignments in cases where more than one filament satisfies the distance criterion defined in Sec. \ref{sec:cosmic_classification}. We have verified that allowing repeated galaxy assignments does not change the main results. This selection yields a final sample of 8513 unique filaments associated with at least one galaxy.

\begin{figure}[h!]
    \centering
    \includegraphics[width=1\linewidth]{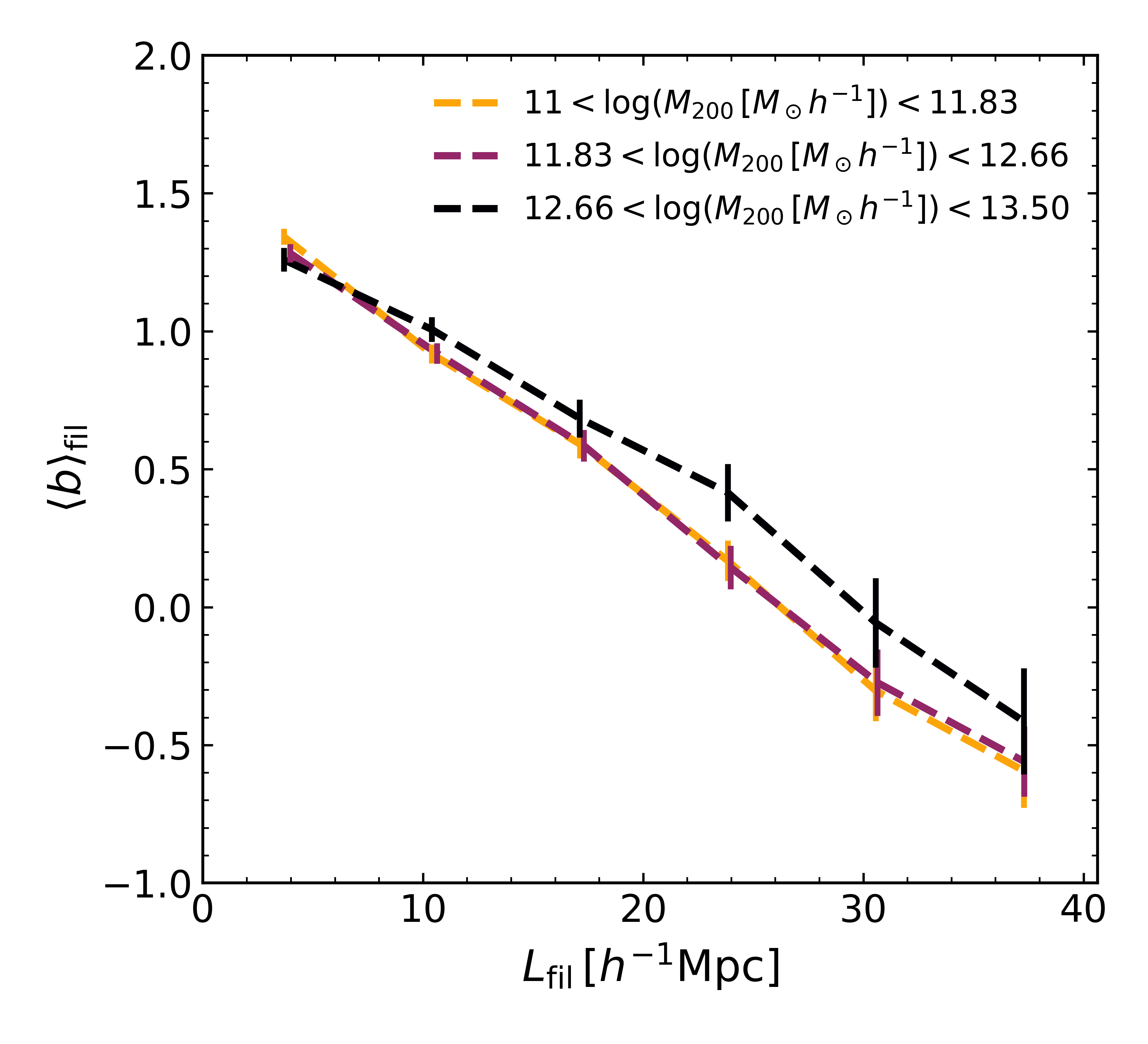}
    \caption{Mean galaxy bias of filaments as a function of filament length, separated by halo mass. The different curves correspond to three halo-mass intervals, as indicated in the legend. The decreasing trend with filament length persists across all mass bins.}
    \label{fig:mean_bias_lfil_massbin}
\end{figure}

Taking advantage of the object-by-object bias estimator, we compute the mean galaxy bias associated with each individual filament as,
\begin{equation}
    \langle b \rangle_{fil} = \frac{1}{N_{G_{fil}}} \: \sum_{i = 1}^{N_{G_{fil}}}b_i \; \; 
\end{equation}
where $N_{G,\mathrm{fil}}$ is the number of galaxies associated with a given filament, and $b_i$ is the individual bias of the $i$-th galaxy.

We first analyze the dependence of this quantity on filament length, as shown in Figure~\ref{fig:mean_bias_fil}. The filament length, $L_{\rm fil}$, is computed as the sum of the lengths of all DisPerSE segments composing each filament. We find a clear decreasing trend, where short filaments reach mean galaxy bias values of $1.4$, while long filaments exhibit lower values, around $-0.5$. This behavior suggests that the large-scale bias of galaxies in filaments may depend on the global properties of the filamentary structure.

One possible interpretation is that shorter filaments are more closely connected to dense regions of the cosmic web, such as nodes or node outskirts \citep{Galarraga2020, GalarragaEspinosa2022}. In this case, their higher mean bias could reflect the influence of nearby massive structures. In contrast, longer filaments may extend across lower-density regions or connect structures through void-like environments, leading to lower mean bias values. This interpretation is consistent with the idea that the individual bias is sensitive not only to the local classification of galaxies as filament members, but also to the broader large-scale environment traced by the filament.

\begin{figure}[h!]
    \centering
    \includegraphics[width=1\linewidth]{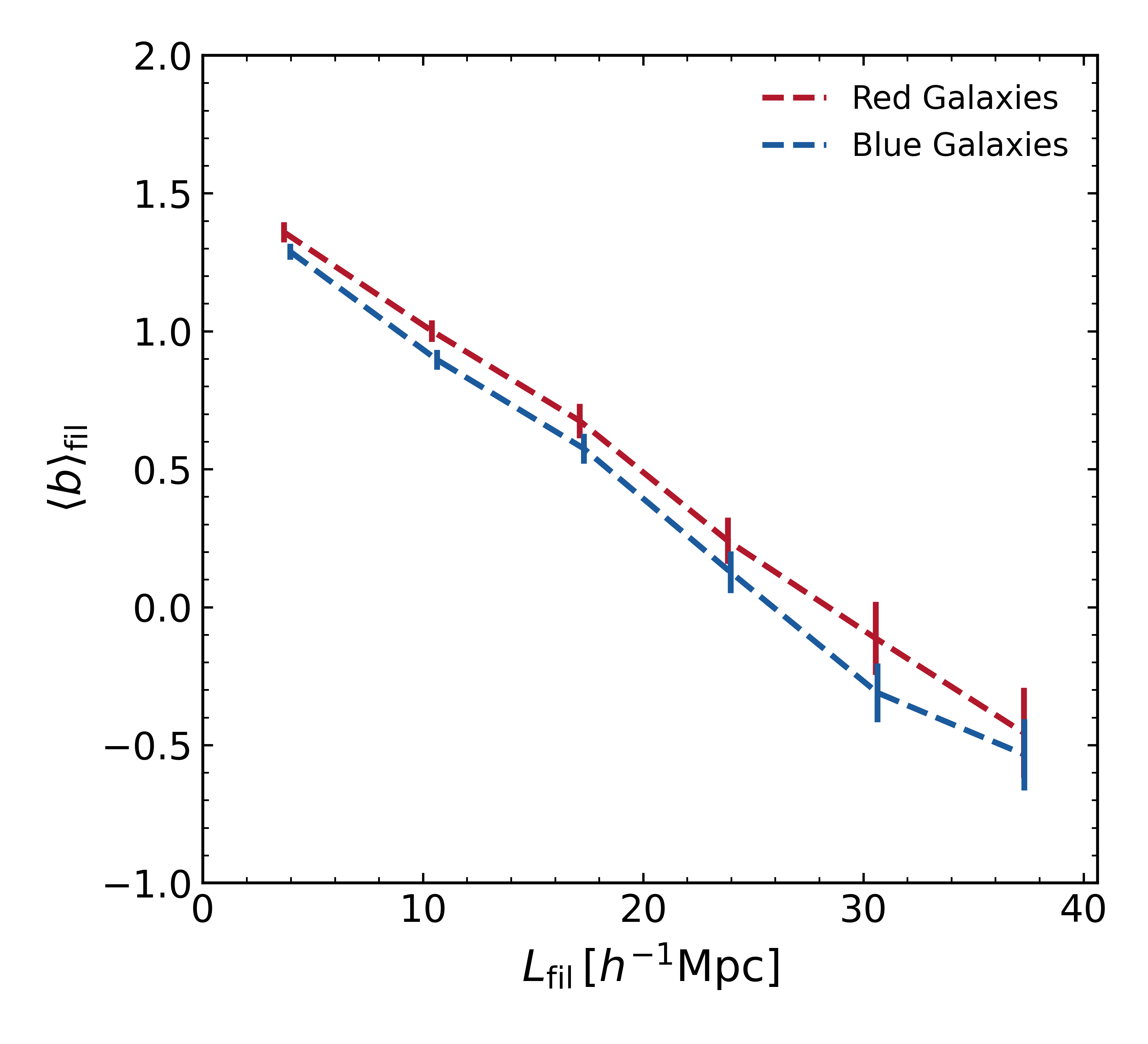}
    \caption{Mean galaxy bias of filaments as a function of filament length, separated by galaxy color. Red and blue curves correspond to red and blue galaxies, respectively. Both populations follow a similar declining trend with filament length, while red galaxies exhibit systematically higher mean bias values.}
    \label{fig:bias_lfil_color}
\end{figure}

However, this trend should be interpreted with caution. The observed relation between mean filament bias and filament length may also be affected by other physical quantities, such as the typical halo mass of galaxies within each filament or the local density around them. To assess this, we recompute the mean bias of each filament by considering only galaxies that satisfy a given selection criterion. We first control for halo mass by dividing the galaxy sample into three different $M_{200}$ intervals, as illustrated in Figure ~\ref{fig:mean_bias_lfil_massbin}. The yellow dashed curve corresponds to the low-mass subsample ($10^{11} < M_{200}\,[h^{-1}{\rm M_{\odot}}] < 10^{11.83}$), the purple dashed curve to the intermediate-mass subsample ($10^{11.83} < M_{200}\,[h^{-1}{\rm M_{\odot}}] < 10^{12.66}$), and the black dashed curve to the high-mass subsample ($10^{12.66} < M_{200}\,[h^{-1}{\rm M_{\odot}}] < 10^{13.50}$). It is clear how the decreasing trend between mean filament bias and filament length remains in all mass thresholds. As expected, galaxies in the higher halo-mass bins exhibit systematically larger bias values, reflecting the well-established tendency of more massive systems to exhibit higher large-scale bias than their lower-mass counterparts. Nevertheless, the overall behavior of the relation remains remarkably similar across the three halo-mass intervals. This suggests that halo mass is not the primary driver of the observed trends, and that additional environmental factors may play a more significant role in shaping galaxy bias within filaments.

We also repeat the analysis by separating galaxies according to color using the same criteria as Sect. \ref{sec:bias_func}. The result is shown in Figure~\ref{fig:bias_lfil_color}. The dependence on filament length is still present for both red and blue galaxies. However, red galaxies exhibit systematically higher mean bias values than blue galaxies across the full range of filament lengths, with differences in bias of up to $\sim 0.20$.

\begin{figure}[h!]
    \centering
    \includegraphics[width=1\linewidth]{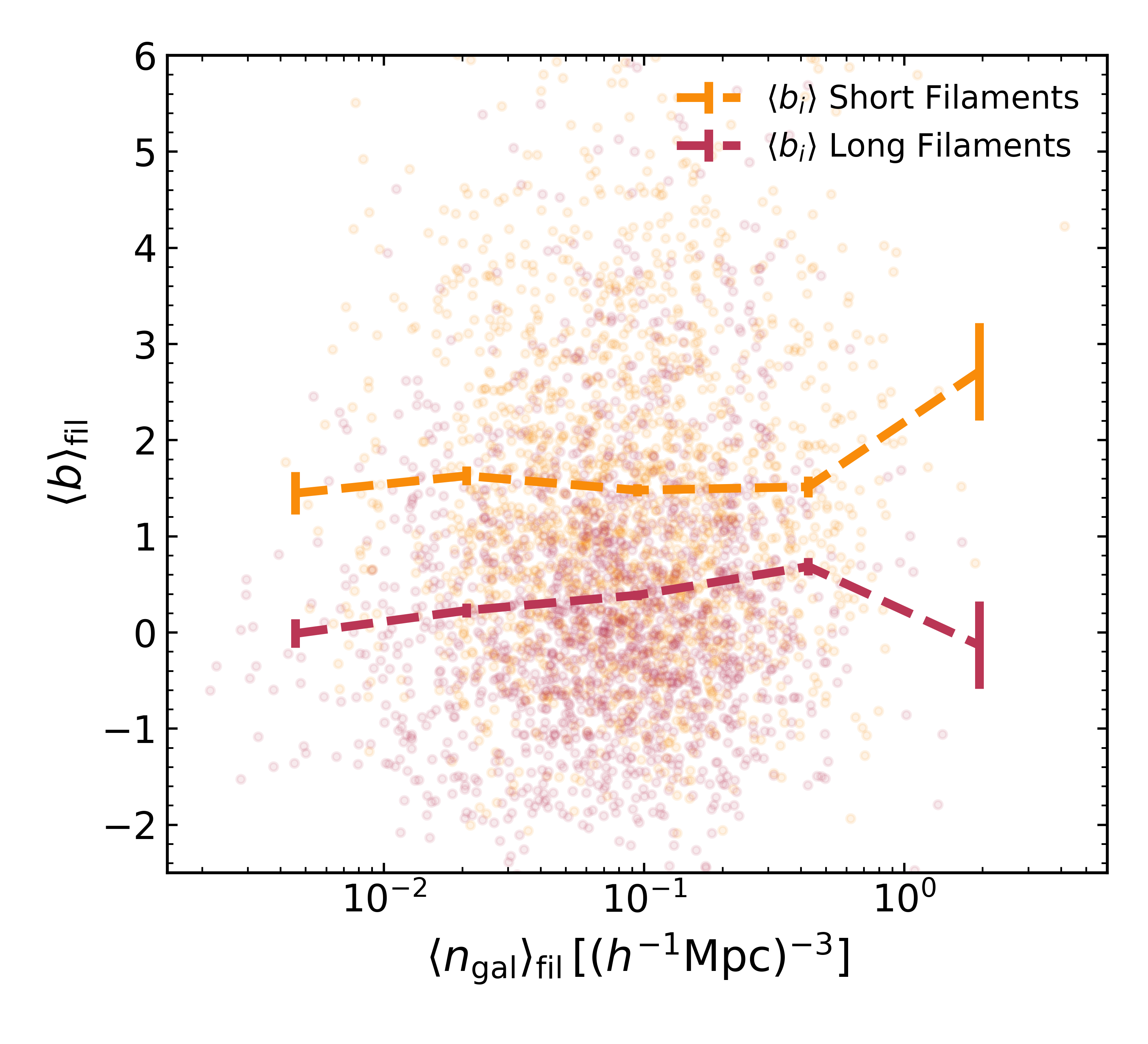}
    \caption{Mean galaxy bias of filaments as a function of filament galaxy density, separated by filament length. Each point represents an individual filament, colored according to whether it belongs to the short- or long-filament population. Dashed lines show the mean trends for each population. The dependence on filament length remains visible at fixed galaxy density.}
    \label{fig:bias_densfil}
\end{figure}

To assess whether the trend with filament length is driven by other filament properties, we study the relation between mean filament bias, $\langle b \rangle_{\rm{fil}}$, and local galaxy density within filaments, $\langle n_{\rm{gal}} \rangle_{\rm{fil}}$. The local galaxy density of each filament is computed as,
\begin{equation}
    \langle n_{\rm{gal}} \rangle_{\rm{fil}} =
\frac{1}{N_{\mathrm{seg}}}
\sum_{i=1}^{N_{\mathrm{seg}}} n_i
\end{equation}
where $N_{\mathrm{seg}}$ is the number of segments in the filament and $n_i$ is the galaxy density associated with the $i$-th segment. The density $n_i$ is computed as the number of galaxies within a cylindrical volume surrounding the $i$-th segment, divided by the corresponding cylinder volume. We adopt a cylinder radius of $1.35\,h^{-1}\mathrm{Mpc}$, corresponding to the outer boundary of the filament-outskirts classification.

Figure~\ref{fig:bias_densfil} shows the relationship between $\langle b \rangle_{\rm{fil}}$ and $\langle n_{\rm{gal}} \rangle_{\rm{fil}}$ for each filament. At first order, we find no significant dependence of the mean filament bias on the average galaxy density of the filament. However, when separating the sample into the $30\%$ shortest (yellow) and $30\%$ longest (red) filaments and measuring the mean values (dashed lines) of each population, the dependence on filament length remains visible at fixed galaxy density. At fixed $\langle n_{\rm gal}\rangle_{\rm fil}$, short filaments exhibit systematically higher mean bias values than long filaments, with differences in bias of typically $\sim 0.8$--$1.5$ across most of the density range. This indicates that the bias-length relation is not primarily driven by variations in filament galaxy density. A more comprehensive analysis of the relationships between $\langle n_{\rm gal} \rangle_{\rm fil}$, $L_{\rm fil}$, and $\langle b \rangle_{\rm fil}$ is provided in Appendix~\ref{app:density_len}. Although additional details are presented there, the main conclusions remain unchanged and are fully consistent with the trends described throughout this subsection.

\subsection{Galaxy bias profile along the filament skeleton}

We now investigate how galaxy bias varies along the filament skeleton, from the saddle point toward the node, and whether this longitudinal dependence changes with filament length. For this purpose, we consider only filaments connected to massive nodes, allowing us to trace how galaxy bias changes when moving along the filament toward or away from these structures. Specifically, we require the densest endpoint of each filament, identified as a CPmax, to lie within $1.5R_{200}$ of the corresponding halo center. We further restrict the analysis to filament segments located outside nodes and node outskirts. This selection yields a sample of 1346 filaments.

To study how galaxy bias varies along filaments, we construct longitudinal bias profiles. For each filament, the individual bias values of its galaxies are normalized by the mean bias of the corresponding filament, $\langle b \rangle_{\rm fil}$. The longitudinal distance of each galaxy is also normalized by the filament length, considering only the filament segments located outside nodes and node outskirts. With this convention, the normalized coordinate ranges from 0 at the saddle point to 1 at the node.

The main results of our analysis are shown in Figure \ref{fig:long_bias}. For short (yellow line; $L_{\rm{fil}}<6.77\; h^{-1} \rm{Mpc}$) and medium (purple line; $6.77 \; h^{-1}\rm{Mpc}$ $\, \leq L_{\rm{fil}}<13.55 \; h^{-1}\rm{Mpc}$) filaments, the normalized galaxy bias, $b_i/\langle b \rangle_{\rm fil}$, remains approximately constant along the filament. This indicates that galaxies in short filaments have bias values similar to the mean bias of their host filament, with no strong dependence on their longitudinal position. In contrast, longer filaments (black line; $13.55 \; h^{-1}\rm{Mpc} \leq $ $L_{\rm{fil}}$) show a clearer longitudinal variation: galaxies close to the saddle point tend to have lower bias values than the filament average, while galaxies closer to the node tend to have higher bias values. For long filaments, the normalized bias increases from $0.85$ near the saddle point to $1.06$ close to the node. This suggests that long filaments trace different environmental conditions along their length, with the regions near nodes being associated with more strongly biased environments.

\section{Interpretation and discussion}\label{sec:disc}

Cosmic filaments are fundamental components of the cosmic web. These extended and anisotropic environments shape the properties of galaxies and halos that inhabit them, as these populations are affected by both the internal conditions of the filamentary structure and the gravitational influence of the dense nodes they bridge \citep[e.g.,][]{kuutma2017,Kraljic2018,Kraljic2019,Galarraga2020,Galarraga2024}. In this work, we employ an object-by-object estimator of large-scale bias \citep{paranjape2018, Balaguera2024, montero2025a, montero2025b, Alfaro2026} to perform a detailed analysis of the galaxy bias within filaments and its relation to global properties such as length and density. Ultimately, this work extends and complements the analysis of galaxy bias within cosmic voids in \cite{montero2025b} and \cite{Alfaro2026}, providing a more complete characterization of galaxy bias across the different environments of the cosmic web.

\begin{figure}
    \centering
    \includegraphics[width=1\linewidth]{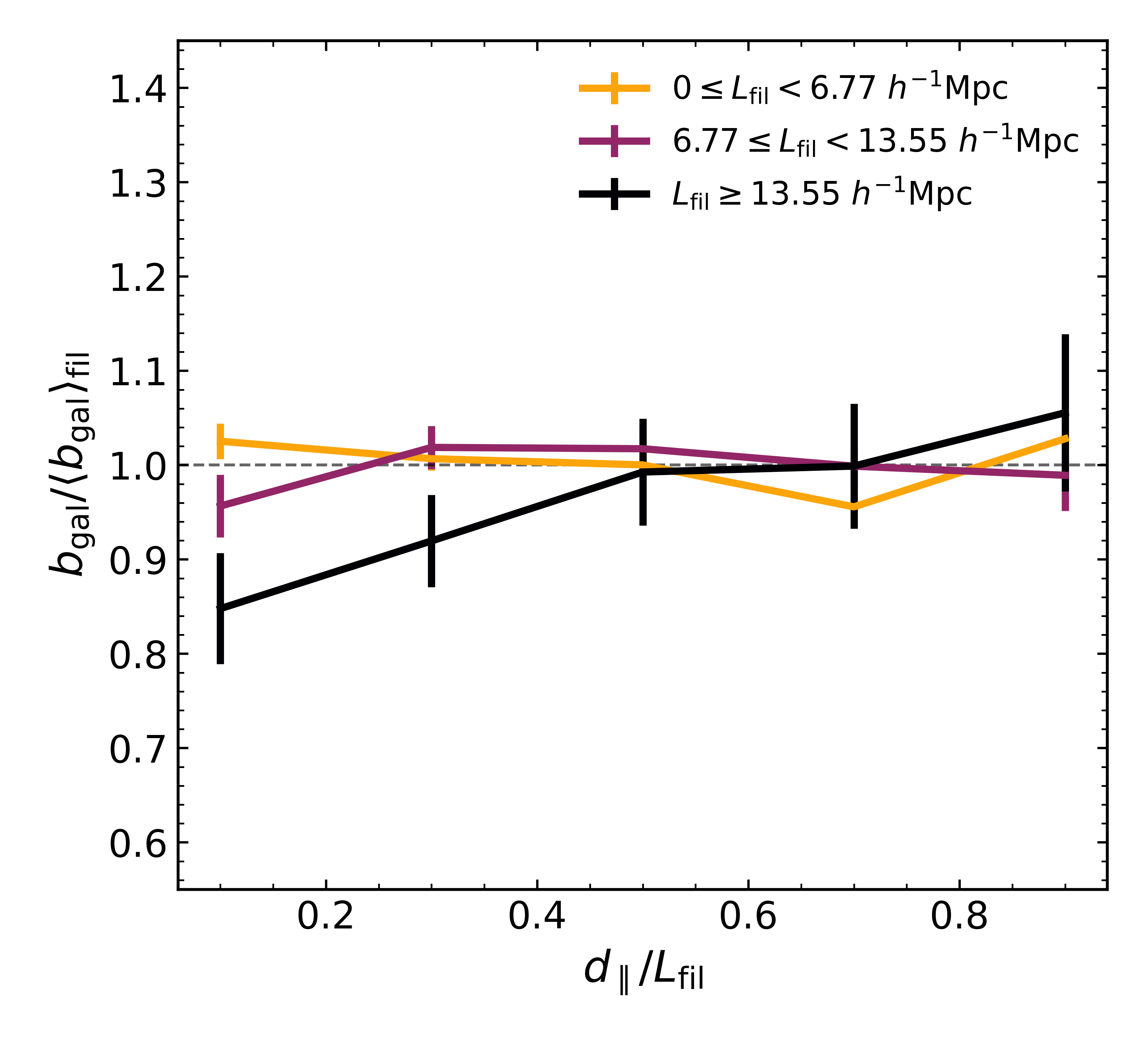}
    \caption{Normalized galaxy bias profile along the filament spine, separated by filament length. The longitudinal distance is normalized by the filament length, with $d_{\parallel}/L_{\rm fil}=0$ corresponding to the saddle point and $d_{\parallel}/L_{\rm fil}=1$ to the node. Short filaments show an approximately flat profile close to unity, while long filaments exhibit a clear increase in bias from the saddle point toward the node.}
    \label{fig:long_bias}
\end{figure}

As shown in Fig. \ref{fig:biasfunc_all_cen_sat}, the strongest departure from the \citet{Tinker2010} fit is found in node outskirts, where the galaxy bias is higher by a factor of up to $\sim4$. This result reveals that the geometric location of galaxies within the cosmic web modulates their large-scale galaxy bias, beyond the primary dependence on halo mass. It is consistent with previous studies showing that, at fixed halo mass, large-scale halo and galaxy bias depend on the cosmic web geometry and on the anisotropy of the surrounding LSS \citep[e.g.,][]{Hahn2009,Borzyszkowski2017,Musso2018,paranjape2018, monterodorta-rodriguez2024, MonteroDorta_Rodriguez2026}.  In particular, our findings align with \cite{paranjape2018}, who also used an object-by-object bias estimator to show that halos in highly anisotropic environments ($\alpha_R > 0.5$) exhibit significantly higher bias than those in isotropic regions. Furthermore, the magnitude of this effect is consistent with \cite{monterodorta-rodriguez2024}, who reported a secondary bias signal in terms of relative bias of $\sim 2$ for galaxies close to nodes at the low-mass end ($M_{200} \leq 10^{12.5}\,h^{-1}{\rm M_\odot}$). This is also supported by its observational counterpart, \cite{MonteroDorta_Rodriguez2026}, where secondary bias was robustly detected in SDSS galaxy groups, showing that groups located near Maxima (nodes) exhibit enhanced relative bias ($b_{relative} > 1$) at fixed halo mass.

This increase in the bias is also found in galaxies hosted by low-mass halos ($M_{200} \leq 10^{12.5} h^{-1}{\rm M_\odot}$) residing in filaments and filament outskirts. The enhancement is particularly pronounced for red galaxies, whose bias function shows a distinct upturn, reaching mean values of $1.4$ and $1.6$, respectively. Such an enhancement for faint red populations is consistent with the findings of \cite{Swanson2008}, who reported that both luminous and dim red galaxies exhibit higher bias values than blue galaxies, reaching relative bias values of $\sim 1.6$. However, this excess is strongly reduced after excluding galaxies located within $5\,h^{-1}{\rm Mpc}$ of a node. The original red-galaxy sample remains above the filtered sample by a factor of up to 1.5 in filaments and 1.4 in filament outskirts.

In both node outskirts and filamentary environments, our results highlight proximity to massive structures as the dominant driver of the environmental secondary bias signal. In these regions, galaxies hosted by low-mass halos effectively inherit the large-scale linear bias of the nearby node, which is primarily determined by halo mass. This inheritance effect, where low-mass halos in the vicinity of massive ones share their large-scale bias properties, has been well documented in the literature, often in connection with populations of splashback or ejected halos \citep[e.g.,][]{Dalal2008,wang2009,tucci2021}. 

Moreover, the upturn in the bias function of red galaxies suggests a connection between the bias signal and two-halo galactic conformity, defined as the correlation between the color or star formation activity of low-mass central galaxies and their neighbors in adjacent halos at separations of several megaparsecs \citep[e.g.,][]{Kauffmann2013,Hearin2015,Hearin2016,Paranjape2015,Lacerna2022,Ayromlou2023}. Our findings align with recent research indicating that this signal is primarily produced by central galaxies residing in the vicinity of massive systems \citep[e.g.,][]{Lacerna2022,Ayromlou2023,Palma2025}. Specifically, \citet{Lacerna2025} have shown that removing galaxies close to massive halos reduces both the conformity and the assembly bias signal, suggesting a close connection between these two phenomena. In this context, the bias excess observed for red galaxies hosted by low-mass halos in filaments and filament outskirts likely reflects this same large-scale environmental influence associated with massive nodes.

In contrast, the bias values measured for galaxies in voids and walls in this work are positive, ranging from $\sim 0.5$ to $1.6$. These values are notably higher than those reported in recent studies \citep[e.g.,][]{Balaguera2024,montero2025b,Alfaro2026}, which use more restrictive void definitions and find negative bias values, from $\sim -0.5$ down to $\sim -2.0$ for the largest underdense structures. This discrepancy likely stems from the adopted cosmic web classification, in which the void-and-wall environment is defined as a residual. Consequently, this category can include galaxies located near the boundaries of denser structures, increasing the mean bias relative to void finders such as \textsc{Sparkling} \citep{ruiz_void_2015,ruiz_into_2019} or \textsc{Popcorn} \citep{paz_2022}, which impose strict integrated underdensity thresholds (e.g., $\Delta_{\rm lim}=-0.9$ in \citealt{montero2025b,Alfaro2026}).

The decrease in mean galaxy bias with filament length can be interpreted in terms of the geometric configuration of the cosmic web. As suggested by \citet{Galarraga2020}, short filaments are intrinsically denser and can act as direct matter bridges between massive structures in high-density environments, whereas long filaments are less dense and may represent the extended skeleton of the cosmic web in less populated regions. Consistently, galaxies in short filaments are expected to reside closer to nodes and, following the interpretation discussed above, effectively inherit their enhanced large-scale bias. In contrast, longer filaments can extend across lower-density regions or void-like environments and thus exhibit bias values characteristic of underdense structures. Importantly, this relation does not appear to be driven by internal galaxy or halo properties. While red galaxies have a higher bias than blue galaxies at fixed filament length, and galaxies in more massive halos also exhibit higher bias values, the decreasing trend with filament length remains consistent across the analyzed mass and color ranges. This result can therefore be interpreted as a secondary bias signal associated with filament properties, in which the galaxy bias depends on the geometry and connectivity of the host structure rather than halo mass. This effect mirrors the findings for cosmic voids in \cite{montero2025b}, where the bias profiles of S-type and R-type voids exhibit different amplitudes modulated by the density of the surrounding environment, a difference that is likewise not driven by halo mass.

In contrast, the mean galaxy bias shows no significant dependence on the average local galaxy density of the filament. Although short filaments tend to be denser \citep{Galarraga2020, GalarragaEspinosa2022}, their enhanced bias persists at fixed filament density, indicating that local galaxy density is not the primary driver of the bias--length relation. Instead, the signal appears to be more closely connected to the large-scale geometry and environment traced by the filament.

Fig.~\ref{fig:long_bias} reveals distinct longitudinal trends depending on filament length. Short filaments exhibit nearly uniform bias profiles, whereas long filaments show a clear increase in bias from the saddle point toward the node. This suggests that long filaments act as transitional environments, connecting lower-density regions near saddles with the denser environments surrounding massive nodes. This interpretation is consistent with previous studies showing that saddle points represent local minima of both galaxy density and stellar mass, with these quantities increasing toward nodes \citep[e.g.,][]{Kraljic2019}. We add to this picture by showing that the individual bias follows a similar longitudinal gradient, increasing from $b_i/\langle b\rangle_{\rm fil}\sim0.85$ near the saddle point to $\sim1.06$ close to the node. This longitudinal variation also suggests an anisotropic dependence of the large-scale bias on the geometry of the cosmic web, although the isotropic estimator employed here cannot explicitly separate its directional components. Thus, the bias profile along long filaments traces the transition from underdense regions toward the high-density nodes of the cosmic web.

\section{Conclusions}

In this work, we used an object-by-object estimate of large-scale linear galaxy bias applied to the TNG300 hydrodynamical simulation to investigate how the bias of galaxies depends on their location within the cosmic web, with particular emphasis on filamentary environments. Filaments were identified using the DisPerSE algorithm. Our main conclusions can be summarized as follows:

\begin{itemize}

    \item Galaxy bias exhibits a clear dependence on cosmic web environment at fixed halo mass, with the strongest enhancement found in node outskirts, where the bias exceeds the \citet{Tinker2010} fit by a factor of up to $\sim4$. A similar enhancement is observed for low-mass red galaxies in filaments and filament outskirts, but is strongly reduced after excluding galaxies within $5 \; h^{-1}{\rm Mpc}$ of a node, with the original samples exceeding the filtered ones by factors of up to 1.5 and 1.4, respectively. Together, these results identify proximity to massive nodes as a dominant driver of environmental secondary bias, consistent with low-mass halos effectively inheriting the large-scale bias properties of nearby massive structures.
    
    \item The mean galaxy bias of individual filaments decreases with filament length, with short filaments reaching values of $1.4$ and long filaments $-0.5$. This trend persists when controlling for halo mass and galaxy color, indicating that it is not primarily driven by internal galaxy or halo properties. We therefore interpret the bias--length relation as an environmental secondary bias signal associated with filament properties.
    
    \item The bias--length relation is not primarily driven by the average local galaxy density of the filament. Although short filaments tend to be denser, they remain systematically more biased than long filaments at fixed filament density. This supports a stronger connection between large-scale galaxy bias and the geometric configuration and connectivity of filaments than the internal population density of the filamentary structure.
    
    \item Short and long filaments exhibit distinct longitudinal bias profiles. Short filaments show an approximately uniform profile, whereas long filaments display a clear increase from the saddle point toward the node, from $b_i/\langle b\rangle_{\rm fil}\sim {\rm 0.85}$ to $\sim {\rm 1.06}$. This gradient suggests that long filaments trace transitional environments from underdense regions toward massive nodes and may reflect an anisotropic dependence of large-scale bias on cosmic web geometry.
\end{itemize}

This study provides a detailed mapping of how galaxy bias is modulated across the filamentary structures of the cosmic web, highlighting their role as fundamental drivers of environmental secondary bias. Together with \cite{montero2025b} and \cite{Alfaro2026}, this study provides a more complete characterization of how galaxy bias is modulated across the cosmic web, from underdense regions such as cosmic voids to filaments and nodes. Characterizing how bias varies across the filamentary network provides valuable insight into the interplay between the growth of LSS and the formation and evolution of galaxies. 

Beyond characterizing the environmental dependence of galaxy bias within filaments, our individually-estimated large-scale bias measurements open several avenues for cosmological application. First, because the estimator assigns a large-scale bias value to each galaxy individually, rather than requiring clustering statistics computed over pre-defined subsamples, it is naturally suited to constructing optimally-weighted or bias-selected tracer populations for multi-tracer analyses \citep{McDonald2009, Seljak2009}, which exploit the correlated sample variance between differently-biased tracers of the same underlying density field to tighten constraints on the growth rate $f\sigma_8$ and on primordial non-Gaussianity $f_{\rm NL}$; this approach has already been applied to real survey data with encouraging results \citep{Abramo2013, Blake2013}, and a bias estimator sensitive to intra-sample variation via filament location could provide a natural, physically-motivated way to split galaxy samples for this purpose. Second, if the large-scale bias exhibits a systematic dependence on filament properties or on a galaxy's position within the filament, this represents a potential systematic for baryon acoustic oscillation (BAO) reconstruction algorithms and redshift-space distortion (RSD) analyses that typically assume a single effective bias per galaxy sample, a point of direct relevance to upcoming Stage-IV surveys such as DESI and Euclid. Third, individually-estimated, spatially-varying bias values could serve as a physically-motivated prior in Bayesian forward-modeling reconstructions of the initial density field, which jointly infer initial conditions and galaxy bias parameters from redshift survey data \citep{Jasche2013, Jasche2015, JascheLavaux2019}; replacing the constant-bias assumption typically adopted in such pipelines with a filament-informed bias field could improve the fidelity of reconstructed density and velocity fields, particularly in filamentary regions where a single effective bias may be a poor approximation.

Although the individual bias estimator used in this work provides a robust object-by-object description of the large-scale bias, it is intrinsically isotropic. It therefore cannot separate the directional components of the bias signal associated with the anisotropic geometry of the cosmic web. A natural extension of this approach would be to use anisotropic estimators to distinguish variations along and across the filamentary skeleton. Furthermore, our results also provide physical motivation for theoretical frameworks such as the Web-Halo Model \citep{Brieden2026}, which explicitly incorporates filaments and sheets as collapsed structures to improve the accuracy of matter clustering predictions. By quantifying how galaxy bias varies with filament length, node proximity, and position along the filament spine, our analysis shows that significant bias differences can arise even within similarly classified cosmic web structures. Finally, extending this analysis to different cosmic web finders, such as NEXUS/NEXUS+, Bisous, or T-Rex, will be essential to assess whether these trends persist across different definitions of the cosmic web, especially in view of future applications to LSS surveys.

%%%%%%%%%%%%%%%%%%%%%%%%%%%%%%%%%%%%%%%%%%%%%%%%%%%%%%%%%%%%%%
\begin{acknowledgements}

CASS and ADMD acknowledge support from the Universidad Técnica Federico Santa María through the Proyecto Interno Regular \texttt{PI\_LIR\_25\_04}. CASS also acknowledges financial support from the Pontificia Universidad Católica de Valparaíso through a maintenance scholarship. ADMD also thanks the ICTP for their hospitality and financial support through the Regular Associates Programme 2022–2027. The authors also acknowledge the AstroGainz science outreach initiative for its support and scientific discussions.
\end{acknowledgements}

% Please note that we have included the references below in
% order to compile the document, but we ask you to:

\bibliographystyle{aa} % style aa.bst
\bibliography{references} % your references Yourfile.bib

% %%%%%%%%%%%%%%%%%%%%%%%%%%%%%%%%%%%%%%%%%%%%%%%%%%%%%%%%%%%%%%
% Example below of non-structurated natbib references  
% To use the v8.3 macros with this form of composition of bibliography,
% the option "bibyear" should be added to the command line
% "\documentclass[bibyear]{aa}".
% %%%%%%%%%%%%%%%%%%%%%%%%%%%%%%%%%%%%%%%%%%%%%%%%%%%%%%%%%%%%%%

% \begin{thebibliography}{}

%   \bibitem[1966]{baker} Baker, N. 1966,
%       in Stellar Evolution,
%       ed.\ R. F. Stein,\& A. G. W. Cameron
%       (Plenum, New York) 333

%    \bibitem[1988]{balluch} Balluch, M. 1988,
%       A\&A, 200, 58

%    \bibitem[1980]{cox} Cox, J. P. 1980,
%       Theory of Stellar Pulsation
%       (Princeton University Press, Princeton) 165

%    \bibitem[1969]{cox69} Cox, A. N.,\& Stewart, J. N. 1969,
%       Academia Nauk, Scientific Information 15, 1

%    \bibitem[1980]{mizuno} Mizuno H. 1980,
%       Prog. Theor. Phys., 64, 544
   
%    \bibitem[1987]{tscharnuter} Tscharnuter W. M. 1987,
%       A\&A, 188, 55
  
%    \bibitem[1992]{terlevich} Terlevich, R. 1992, in ASP Conf. Ser. 31,
%       Relationships between Active Galactic Nuclei and Starburst Galaxies,
%       ed. A. V. Filippenko, 13

%    \bibitem[1980a]{yorke80a} Yorke, H. W. 1980a,
%       A\&A, 86, 286

%    \bibitem[1997]{zheng} Zheng, W., Davidsen, A. F., Tytler, D. \& Kriss, G. A.
%       1997, preprint
% \end{thebibliography}

%%%%%%%%%%%%%%%%%%%%%%%%%%%%%%%%%%%%%%%%%%%%%%%%%%%%%%%%%%%%%%%
% Appendices must be placed after   \end{thebibliography}
% They will be placed automatically on a new page.
%%%%%%%%%%%%%%%%%%%%%%%%%%%%%%%%%%%%%%%%%%%%%%%%%%%%%%%%%%%%%%%
\begin{appendix}

%---------------------------------------------

% Force figure numbering for Appendix A
\renewcommand{\thefigure}{A.\arabic{figure}}
\setcounter{figure}{0}

\begin{strip}
\centering
\includegraphics[width=1.03\textwidth]{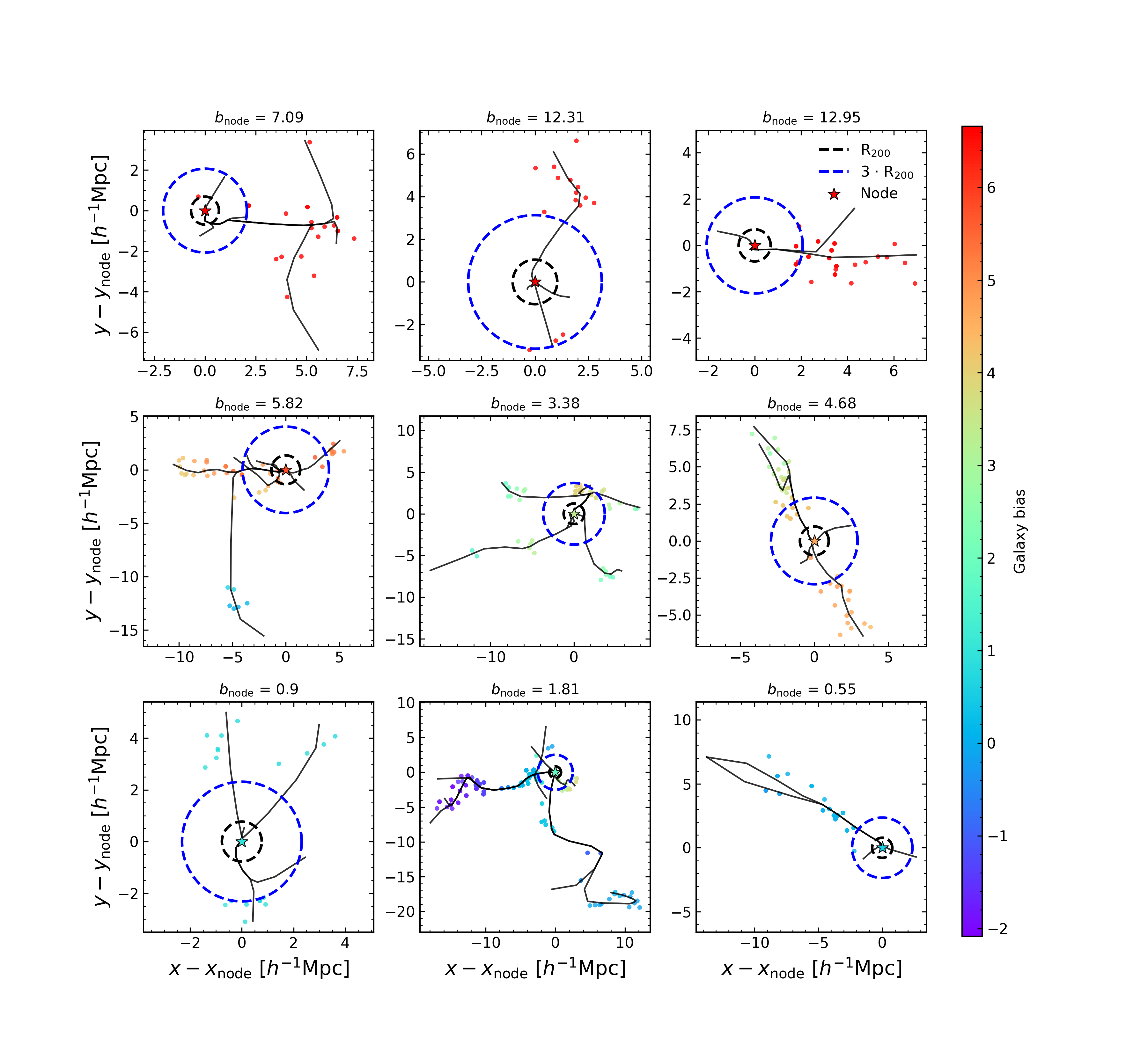}
\captionof{figure}{
Examples of filamentary structures connected to massive nodes. Each panel is centered on a node, whose position is marked by a red star. Black curves show the DisPerSE filament skeleton, while points correspond to galaxies associated with the filamentary structure and are colored according to their individual galaxy bias. The black and blue dashed circles indicate $R_{200}$ and $3R_{200}$ around the node, respectively. The title of each panel shows the bias value of the corresponding node. These examples illustrate how galaxies located along filaments can exhibit a broad range of bias values depending on their position relative to the node and the surrounding large-scale environment.}
\label{fig:app_bias_filaments}
\vspace{0.5cm}
\end{strip}

\section{Visualize Bias in Filaments}
\label{app:bias_fil}

In this appendix, we provide a visual illustration of the spatial distribution of individual galaxy bias in filamentary environments connected to massive nodes. Figure~\ref{fig:app_bias_filaments} shows a set of representative examples centered on massive nodes. In each panel, the DisPerSE filament skeleton is shown together with the galaxies associated with the filamentary structure, colored according to their individual bias values.

% Force figure numbering for Appendix B
\renewcommand{\thefigure}{B.\arabic{figure}}
\setcounter{figure}{0}

\begin{figure}[!htbp]
    \centering
    \includegraphics[width=1.03\linewidth]{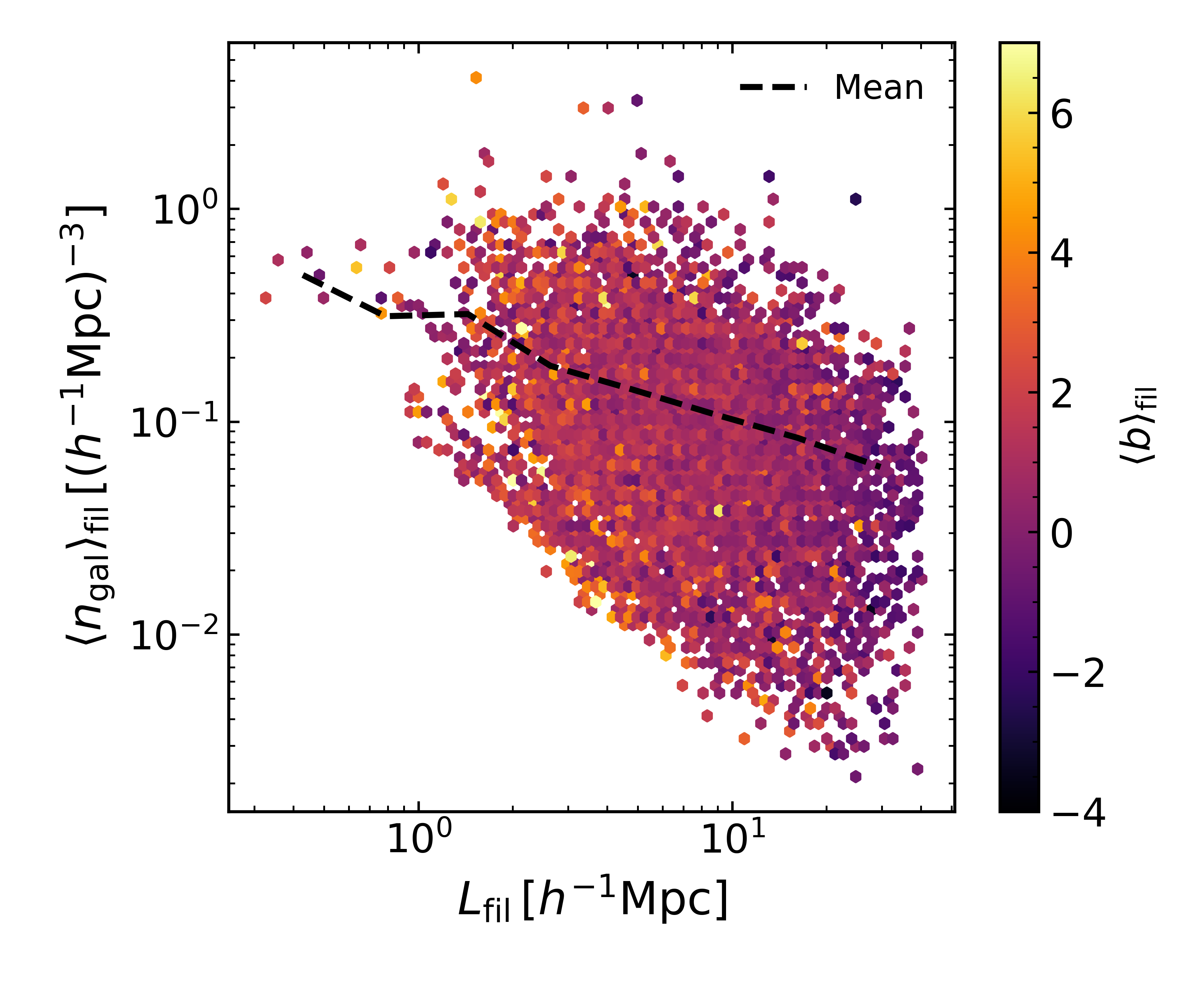}
    \caption{
    Average local galaxy density of individual filaments as a function of filament length. Each point represents a single filament, colored by its mean galaxy bias. The black dashed curve shows the mean trend. Shorter filaments tend to be denser and more strongly biased, while longer filaments are generally associated with lower galaxy densities and lower mean bias values.
    }
    \label{fig:app_density_length}
\end{figure}

These examples illustrate that galaxies located within the same filamentary system can exhibit a broad range of bias values. In particular, high-bias galaxies are often found close to the node or along filamentary branches connected to it, while lower-bias galaxies can appear farther from the densest region. This behavior highlights that the individual galaxy bias is sensitive not only to the local filament classification, but also to the broader large-scale environment surrounding each galaxy. Therefore, even within a single cosmic web component, galaxies can trace different large-scale bias regimes depending on their position relative to massive nodes.

\section{Density--Length Relation}
\label{app:density_len}

In Sect.~4.1, we showed that the mean galaxy bias of individual filaments decreases with filament length and that this trend is not primarily driven by the average local galaxy density of the filament. To further explore the connection between these quantities, Figure~\ref{fig:app_density_length} shows the relation between the local filament galaxy density, $\langle n_{\rm gal}\rangle_{\rm fil}$, and filament length, $L_{\rm fil}$, for the full filament sample.

The figure shows that shorter filaments tend to have higher average galaxy densities, whereas longer filaments are generally associated with lower densities. This trend is consistent with the interpretation that short filaments preferentially reside in denser regions of the cosmic web, often close to massive nodes, while longer filaments can extend across less populated environments. The points are colored by the mean galaxy bias of each filament, $\langle b\rangle_{\rm fil}$, showing that high-bias systems are preferentially associated with the short filament population.

However, this relation does not imply that the local filament galaxy density is the main driver of the bias--length trend. As discussed in Sect.~4.1, the dependence of mean galaxy bias on filament length remains visible even when filaments are compared at fixed local galaxy density. This supports the idea that the large-scale bias signal is more closely connected to the global geometry and connectivity of the filamentary structure than to its internal galaxy density alone.

%---------------------------------------------

\end{appendix}
\end{document}